\documentclass[12pt]{article}

\pdfoutput=1
\usepackage[margin=1.5in]{geometry}
\usepackage{setspace}
\usepackage{graphicx}
\usepackage{amsmath}
\usepackage{amssymb}
\usepackage{bbm}
\usepackage{authblk}
\usepackage{amsfonts}
\usepackage{amsmath}
\usepackage{amssymb}
\usepackage[T1]{fontenc}
\usepackage{hyperref}
\usepackage[utf8]{inputenc}
\usepackage{orcidlink}
\usepackage{physics}
\usepackage{times}
\usepackage{xcolor}

\title{Quantum hypergraph states: a review}
\author[1]{Vinícius Salem}
\affil[1]{Departamento de F\'{i}sica Te\'{o}rica, At\'{o}mica y \'{O}ptica,
 \\Universidad de Valladolid, 
 47011 Valladolid, Spain}

\date{}

\begin{document}

\maketitle


\hypersetup{
    colorlinks=true,linkcolor=blue,citecolor=blue,
    filecolor=blue,urlcolor=blue,breaklinks=true
}

 \newcommand{\revision}[1]{{\color{red}{#1}}}
 \newcommand{\inclusion}[1]{{\color{blue}{#1}}}
 \newcommand{\doubt}[1]{{\color{purple}{#1}}}


\newcommand{\orcidvinicius}{\orcidlink{0000-0002-1768-8783}}
\newcommand{\orcidalison}{\orcidlink{0000-0003-3552-8780}}
\newcommand{\orcidfabiano}{\orcidlink{0000-0002-8882-2169}}

\newtheorem{theorem}{Theorem}
\newtheorem{corollary}{Corollary}[theorem]
\newtheorem{lemma}[theorem]{Lemma}
\newtheorem{definition}{Definition}[section]
\newtheorem{proposition}[theorem]{Proposition}

 \def\blue{\textcolor{blue}}
 \def\red{\textcolor{red}}
 \def\cyan{\textcolor{cyan}}
 \def\magenta{\textcolor{magenta}}

 \def\be{\begin{equation}}
 \def\ee{\end{equation}}

 \def\bc{\begin{center}}
 \def\ec{\end{center}}
\def\bal{\begin{align}}
\def\eal{\end{align}}
\newcommand{\avg}[1]{\langle{#1}\rangle}
\newcommand{\Avg}[1]{\left\langle{#1}\right\rangle}






\begin{abstract}
Quantum hypergraph states emerged in the literature as a generalization of graph states, and since then, considerable progress has been made toward implementing this class of genuine multipartite entangled states for quantum information and computation. Here, we review the definition of hypergraph states and their main applications so far, both in discrete-variable and continuous-variable quantum information. 
\end{abstract}


\section{Introduction}

Perhaps not since the long-running controversy about the fundamental nature of light has the history of physics seen such a debate as the discovery of a “suspicious phenomenon of action at a distance”, to quote the famous phrase that described entanglement in the pioneering work of Einstein, Podolsky, and Rosen \cite{PR.47.777.1935, walls}. 
Shortly after the publication of EPR, the term ``entanglement'' was coined in the literature by Erwin Schrödinger, originally from the German \textit{``verschränkung''}, which is related to the noun \textit{``schranke''}, meaning ``boundary'' or ``barrier'', like the barriers used at railway crossings \cite{Schrödinger_1935}. The idea brought up by this noun can be related to our habit of crossing our arms \textit{``die Arme verschränken''}, so that both are so close together that it is impossible to move one arm without affecting the other. This same idea was transposed by Schrödinger to two particles, deriving the term \textit{``quantenverschränkung''} \cite{gisin2014quantenverschrankung}.
Although Einstein didn't live to see this epic unfold, the formal definition of the phenomenon of quantum entanglement matured in subsequent decades, mainly in John S. Bell's seminal article in 1964 \cite{Physics.1.195.1964}, which assumed locality as a hypothesis \textit{a priori}, in line with EPR's article, and showed that it is possible to express correlations between the states of the composite system with a given upper limit for any complete theory of quantum mechanics, demonstrating the unreasonableness of a local system with hidden variables, one of the forms defended by EPR.
Although a definitive, formal definition of quantum entanglement remains challenging \cite{BENATTI20201}, due to its crucial role in quantum communications and protocols as a resource, a state is often said to be entangled if it serves as a resource for a nonclassical task \cite{masanes2008useful}.
Thus, entanglement is consolidated as the leading resource for the development of quantum protocols, namely: quantum codes \cite{clark2005efficient}, quantum cryptography \cite{RevModPhys.74.145}, quantum error correction \cite{lidar2013quantum}, quantum internet \cite{PhysRevLett.124.210501}, and quantum metrology \cite{PhysRevLett.96.010401}. The manipulation of entanglement itself is a frequent target of studies to increase or decrease the entanglement of a given quantum state, so it is essential to discern and characterize the various classes of multipartite states as completely as possible. In addition, efforts have been made to model new hierarchies that enable more efficient protocols in quantum computing \cite{PhysRevA.98.062328,Book.2010.MichaelA.Nielsen}.

Hypergraph states and graph states are one of the best examples of multiparticle maximally entangled states, a property often desired for quantum resources \cite{enriquez2016maximally}. Among the graph states, the \textit{cluster} states are beneficial in measurement-based quantum computing \cite{PhysRevA.73.033818,PRA.69.062311.2004}, where entanglement is used as a resource for destructive measurements and the formation of logic gates \cite{PRL.86.5188.2001a,walther2005experimental}.
Hypergraph states appeared in the literature in 2013 as generalizations of graph states, based on the mathematical work of Berge in the 70s  \cite{Book.2001.Berge, graphsandhypergraphs}, and have attracted significant interest for generalizing graph state protocols \cite{PRA.87.022311.2013}. Every graph is a particular case of a hypergraph, so several properties of graph states are amenable to generalization to hypergraph states \cite{clark2005efficient,thomas2022efficient,PhysRevLett.97.143601}. Due to their more general nature, hypergraph states have garnered considerable interest in recent literature, both in purely theoretical contexts and in implementations that utilize them as resources in quantum protocols and quantum networks \cite{NJP.15.113022.2013,huang2024demonstration,yang2021representations}. Physically, hypergraph states may appear in a
spin gases model \cite{PhysRevLett.95.180502}, where a classical gas of particles carrying a qubit degree of freedom and initially set to the superposition state $\ket{+}$ undergoes dynamics during the particles' collision, which effectively induces a controlled phase
gate onto the participating qubits. Following an example given in \cite{PhysRevA.90.042308}, just as a two-particle collision might induce a two-qubit phase gate, which corresponds to drawing an edge between the two vertices representing the qubits, the analogy for a $k-$body interaction is natural, creating $k-$qubit phase gates, graphically represented by a hyperedge in the hypergraph.   

In this work, we provide an up-to-date analysis of hypergraph states and their applications in quantum computation, focusing on their multipartite entanglement, nonlocality, and practical implementation. We also review hypergraph states in the continuous-variables scenario, comparing them with the usual discrete-variable scenario, and propose open problems, mainly generalizations of previous work on graph states. The review is organized as follows. In Section \ref{sec1}, we provide a brief introduction to quantum entanglement, focusing on multiparticle entanglement. In Section \ref{sec3}, we provide all the necessary definitions for understanding hypergraphs as mathematical entities and for hypergraphs as representing quantum states, and briefly discuss their generalizations for qudit states, mixed states, and LME states. Due to its importance in many applications, we dedicate Section \ref{sec4} exclusively to the definition of the stabilizer formalism, magic, and its possible generalizations. In sections \ref{sec5} and \ref{sec6}, we provide a concise review of local operations and equivalence classes in hypergraph states. Section \ref{sec7} introduces the reader to local complementation on hypergraphs, and Section \ref{sec8} presents the main results in the literature on Bell inequalities using hypergraph states. Finally, in Section \ref{sec9} we present a concise review of hypergraph states in continuous variables, finishing the review with an overview of the field, concluding remarks, and open works for interested researchers.

\tableofcontents

\section{Quantum Entanglement}
\label{sec1}
\subsection{Bipartite Entanglement}
The usual first approach to quantum entanglement is bipartite entanglement. Assuming that a separabe pure state $\ket{\psi}$ is shared between two labs $A$ and $B$ with local pure states
$\ket{\psi_A} \in \mathcal{H_A}$ and  $\ket{\psi_B} \in \mathcal{H_B}$, we have the following definition.
\vspace{0.1cm}

\begin{definition}
A bipartite pure state $\ket{\psi}\in \mathcal{H_A} \otimes \mathcal{H_B} $ is a product of states if it can be written as the tensor product of two pure states $\ket{\psi_A} \in \mathcal{H_A}$ e $\ket{\psi_B} \in \mathcal{H_B}$:    
\end{definition}
\begin{equation}
\ket{\psi}=\ket{\psi_A}\otimes\ket{\psi_B},
\end{equation}
\noindent otherwise, the state is nonseparable, and therefore, entangled. Examples of separable states are states such as $\ket{\psi}=\ket{10}$, $\ket{\chi}=(\ket{0}+\ket{1})+(\ket{1}+\ket{0})$ and $\ket{\phi}=\ket{00}+\ket{01}+\ket{10}+\ket{11}$. Two systems are in the same pure entangled state if both are correlated and no other system is correlated with them. 
Also, for any two-part pure entangled state $\ket{\psi}_{12}\in \mathcal{H}_1 \otimes \mathcal{H}_2$ there is an orthonormal Schmidt basis $\{\ket{\phi_i},\ket{\chi_i}\}$ s.t.
\begin{equation}
    \ket{\Psi}_{12}=\sum_i^d c_i \ket{\phi_i}\otimes\ket{\chi_i}
\end{equation}
\noindent whose summation is on the smaller dimension of the bipartite system $d=\min(d_1,d_2)$. 

\begin{definition}
A quantum state $\ket{\Psi}^{ab}$ is maximally entangled if and only if the reduced state at one qubit is maximally mixed, i.e., $\Tr_a{\ketbra{\Psi}{\Psi}}=\frac{1}{2}\textbf{I}_b$. If the maximal entanglement is present across all bipartitions, the state is said to be an absolutely maximally entangled (AME) state \cite{helwig2013absolutely}.
\end{definition}

AME states are useful for quantum secret sharing and teleportation, for example \cite{PhysRevA.86.052335}. For sufficiently large local dimensions, AME states always exist. The  GHZ (Greenberger-Horne-Zeilinger) state is the most famous example of a state with multiparticle entanglement and is a well-known AME state for three qubits \cite{carteret1999multiparticle}. Graph states can represent AME states for five and six qubits, corresponding to the stabilizer codes for quantum error correction \cite{PhysRevA.69.052330,PhysRevLett.118.200502,fujii2015quantum}. The correspondence between hypergraph states and AME states remains an open area of research. 

Examples of maximally entangled states in the space $\mathcal{H}_1 \otimes \mathcal{H}_2$ are the two-qubit pure entangled states Bell states, given by: 
\begin{subequations}
\begin{align}
\ket{\Phi}^{+} =\frac{1}{\sqrt{2}}(\ket{0}_{A} \otimes \ket{0}_{B} + \ket{1}_{A} \otimes \ket{1}_{B}) 
\\
\ket{\Phi}^{-} =\frac{1}{\sqrt{2}}(\ket{0}_{A} \otimes \ket{0}_{B} - \ket{1}_{A} \otimes \ket{1}_{B}) 
\\
\ket{\Psi}^{+} =\frac{1}{\sqrt{2}}(\ket{0}_{A} \otimes \ket{1}_{B} + \ket{1}_{A} \otimes \ket{0}_{B}) 
\\
\ket{\Psi}^{-} =\frac{1}{\sqrt{2}}(\ket{0}_{A} \otimes \ket{1}_{B} - \ket{1}_{A} \otimes \ket{0}_{B}), 
\end{align}
\end{subequations}
    \noindent quite used on applications in bipartite entanglement \cite{weinfurter1994experimental}. The concept can be extended to mixed states, as follows.
\vspace{0.1cm}
\begin{definition}
    A mixed state $\rho$ of a composed system with two parts $A$ and $B$ is separable, i.e., non-entangled, if it can be written as a convex combination of product states:
\end{definition}
\begin{equation}
    \rho_{sep} = \sum_i p_i \rho_i^A \otimes \rho_i^B,
\end{equation}

\noindent where $p_i$ are the probabilities associated to the states of the subsystems $\rho_i^A$ and $\rho_i^B$.

Otherwise, the state is entangled. A state is mixed in the sense that it is not possible to know precisely the preparation of the system, but only a statistical ensemble of possible outcomes. In this case, another observer may fully understand the system preparation and describe it as a pure state. Nonetheless, if the mixed state is entangled, the very presence of entanglement forbids any observer from having complete knowledge of the state subsystems, making it impossible to describe it as a pure state \cite{RMP.81.865.2009}. Since the set of separable states is convex, the combination of two or more separable states is still a separable state.

\subsection{Multiparticle entanglement}

Composite systems involving more than two parts must be presented separately, considering their different properties, usually making the quantification
of entanglement more complicated (\cite{m2019tripartite,PRL.106.190502.2011}).
Generalizing the states product definition for $N$-parts, we have the follow: 
\vspace{0.1cm}

\begin{definition}
    A $N$-partite pure state $\ket{\psi}$ is a product state if it can be written as a tensor product of $N$ pure local states:
\end{definition}
\begin{equation} 
    \ket{\psi}=\ket{\psi_1}\otimes\ket{\psi_2}\otimes...\otimes\ket{\psi_N},
\end{equation}
\noindent Otherwise, the state is entangled. The generalization for mixed states is given below.
\begin{equation}
    \rho = \sum_i p_i \rho_i^A\otimes \rho_i^B\ \otimes ... \otimes \rho_i^N\,
\end{equation}
\noindent where $\rho = p_i …$ are the probabilities associated to the states of the subsystems $\rho_i^A$,...,$\rho_i^N$. 

\begin{definition}
    If a pure multipartite state $\ket{\psi}$ cannot be written as a tensor product of any of its subsystems, including any bipartition, the state is said to be genuinely multipartite entangled (GME). 
\end{definition} 

Hypergraph states are examples of GME states.  
This makes clear the fact that GME is a condition more general than bipartite entanglement.
It is valid to emphasize that any separable state is necessarily a PPT state, and therefore any biseparable state is a PPT mixture. Thus, if a state is not a PPT mixture, it is not biseparable and therefore is a GME state \cite{RMP.81.865.2009,PhysRevLett.77.1413} 

A particular quantum state that presents GME is the GHZ state and $W$ states, written as:
\begin{equation}
    \ket{GHZ}=\frac{\ket{000}+\ket{111}}{\sqrt{2}},
    \label{ghz}
\end{equation}
\begin{equation}
    \ket{W}=\frac{\ket{001}+\ket{010}+\ket{100}}{\sqrt{3}}.
    \label{W}
\end{equation}
As we shall see, hypergraphs can represent the GHZ state, which is used to achieve maximal violation of Bell inequalities, and is an essential resource for quantum communication in multipartite systems through secret sharing, for example, \cite{PRA.59.1829.1999,PhysRevLett.104.210501,PhysRevLett.94.060501}. The $W$ states, on the other hand, present larger robustness in relation to loss of particles, since GHZ states lose the entanglement completely when one of the qubits is lost, an appealing property for systems under too much noise. 
\section{Definitions}
\label{sec3}
\subsection{Basic hypergraph theory}

Before introducing the physical association between quantum states and hypergraphs, the mathematical definition of a hypergraph is presented.
Unlike spectral graph theory, which is well-established in the field, spectral theory for hypergraph states remains under development \cite{banerjee2021spectrum,wang2018development,cooper2012spectra,bretto2013applications,berge1990optimisation}. 
\begin{definition}
    Be $V=\{v_1,v_2,...v_n\}$ a finite set. A hypergraph $H$ over $V$ is a Sperner family $\mathcal{H}=\{E_1,E_2,...,E_m\}$ of subsets of $V$ such that \cite{Book.2001.Berge,voloshin2009introduction}
\end{definition} 
\begin{equation}
    E_i \neq \emptyset, \hspace{1cm} i=1,2,...,n
\end{equation}
\noindent and
\begin{equation}
    \bigcup_{i=1}^m E_i=V. 
\end{equation}
The Sperner family of simple hypergraphs (i.e., those without loops) characterizes hypergraphs whose elements of the set  $\mathcal{H}=\{E_1,E_2,...,E_m\}$ obey the relation $ E_i\subset E_j, i=j$. As in the usual definitions for graphs, the elements  $v_1,v_2,...,v_n$ of $V$ correspond to the vertices, but the sets $E_1,E_2,...,E_m$ constitute the hyperedges of the hypergraph. Since they are defined by sets, hypergraphs are algebraic structures, meaning that the exact form that we draw the hyperedge is irrelevant for the definition, as long as it contains the proper vertices. An illustration of a hypergraph is shown in the figure below.
A multigraph that contains $ \textit{loops}$ and multiple edges is a hypergraph in which each edge has cardinality 2. The combinatory possibilities for a hypergraph are considerably higher than those of a graph. While there are $2^{N(N-1)/2}$ possible graphs for $ N$ vertices, for the same number of vertices there are $2^{2^N}$ possible hypergraphs. This implies the fact that hypergraphs are complex in the sense of Kolmogorov complexity \cite{JPA.47.335303.2014,nannen2010short}. 

Two vertices of a hypergraph are adjacent if the same hyperedge connects them. The degree of a vertex is defined as the number of adjacent vertices to it, i.e., the number of hyperedges connected to this vertex. A hypergraph is considered regular if all the vertices have the same degree \cite{Book.2013.Bretto}. The concept of uniformity can be extended to hypergraphs as defined below. 
\begin{definition}
  A $k$-uniform hypergraph $(V, E)$ is a hypergraph such that $|e| = k$ for all $e \in E$.
\end{definition}

So that all hyperedges have cardinality $k=|e|$. The uniformity condition imposes an additional structure on the hypergraph. Another essential concept is the hypergraph completeness, defined as follows.
\begin{definition}
  A complete $k$-uniform hypergraph $(V, E)$ is a $k$-uniform hypergraph in which every subset of $V$ with $k$ elements is a hyperedge.
\end{definition}
Since the spectral theory of hypergraphs is still in development and it is a topic for debate, it is possible to find different approaches in providing a proper definition of a matrix or a tensor of adjacency of a hypergraph \cite{ouvrard2020hypergraphs,banerjee2021spectrum,wang2018development,cooper2012spectra}. The adjacency matrix of a hypergraph can be drawn from the concept of the adjacency of a graph, where a square matrix indicates whether pairs of vertices are adjacent. For hypergraphs, we define the adjacency matrix as follows. 

\begin{definition}
    The adjacency matrix $A(a_{ij})$ for a hypergraph is given by
\begin{equation}
     a_{ij} =\begin{cases}
    \omega_{e_k}, & $if$ \hspace{0.1cm} v_i \in e_j\\
        0, & $otherwise$
\end{cases}
\end{equation}
\noindent where $\omega_{e_k}$ $\in$ $\mathcal{R}$ are the weights on each hyperedge, considered equal to unity for unweighted hypergraphs.
\end{definition}
 Thus, the adjacency notion still captures binary relationships modelled by an adjacency matrix, and this definition always coincides with the one for complete graphs. A more general idea of adjacency that adequately describes the p-adic relationship in hypergraphs is called the e-adjacency tensor, and differentiates between levels of adjacency among the hyperedges that contain $p\geq 1$ vertices, and therefore is no longer a matrix \cite{OUVRARD201871}. 

\begin{definition}
    Let $\mathcal{H}=(V,E)$ be a hypergraph. Let $e \in E$. The vertices constituting $e$ are said $e-$adjacent vertices \cite{PhysRevE.86.056111}. 
\end{definition}
As in graph theory, for hypergraphs one has the concept of regularity, as follows.
\begin{definition}
     Let $\mathcal{H}=(V,E)$ be a hypergraph. The hypergraph is said to be d-regular if every vertex has degree $d$, i.e., if it's contained in exactly $d$ hyperedges. 
\end{definition}
While the concept of regularity refers to the degree of its vertices, the uniformity refers to the hyperedges:
\begin{definition}
    For a k-uniform hypergraph, its normalized k-adjacency tensor is given by 
\begin{equation}
    \bar a_{i{_1}...i_k} = \begin{cases} \frac{1}{(k-1)!} \prod_{1\leq j \leq k} \frac{1}{\sqrt[k]{d_i{_j}}} \hspace{0.1cm} & $if$ \hspace{0.1cm} v_{i{_1}}...v_{i{_k}} \in $E$
      \\ 0,& $otherwise$
    \end{cases}
\end{equation}
\noindent where $E={e_1,e_2...e_p}$ is the family of hyperedges. 
\end{definition}

Naturally, if $\mathcal{H}$ is a $k-$uniform hypergraph, then the concepts of $k-$adjacency and $e-$adjacency are equivalent. More general definitions can be found in \cite{banerjee2017spectra}.
%
%
Hypergraphs are also useful mathematical tools for representing complex systems in networks where the 2-uniform graph is insufficient to provide complete information about the structure of the system \cite{estrada2006subgraph,PhysRevE.107.024316}. In category theory, the collection of hypergraphs is a category with hypergraph homomorphisms as morphisms \cite{fong2019hypergraph}.
\begin{figure}[t]
  \centering
  \includegraphics[width=12cm]{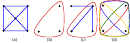}
  \caption{Hypergraphs are generalizations of graphs. (a) A regular graph with 5 vertices, (b) A single-hyperedge hypergraph, (c) the same trivial hypergraph combined with a star graph, (d) a 4-uniform hypergraph representing a GHZ state.}
  \label{Fig1}
\end{figure}
\subsection{Quantum hypergraph states}
First, we present the usual definition for quantum hypergraph states as pure states for qubits. Later, we shall see generalizations for qudit and mixed quantum hypergraph states.
Given a hypergraph $H$ on $n$ vertices, the hypergraph state, denoted by
$\ket{H}$, is obtained in the following way.
First, we assign a qubit for each vertex of the hypergraph in the state
$\ket{+}=(\ket{0}+ \ket{1})/\sqrt{2}$, in such a way that the initial
state is $\ket{+}^{\otimes n}$.
Then, we perform the application of a non-local multiqubit phase gate
$C_e$ acting on the Hilbert spaces associated with vertices $v_i\in e$,
which is a $2^{|e|} \times 2^{|e|}$ diagonal matrix given by
\begin{equation}
    C_e= \mathbbm{1} -2\ketbra{1 \ldots 1},
\end{equation}
\noindent where $\mathbbm{1}$ is the identity matrix. 

This is a generalization of an Ising interaction type, which originated the construction of graph states \cite{Inproceedings.2006.Hein,GNATENKO2021127248}.
Thus, a hypergraph state $\ket{H}$ is a pure quantum state defined as
\cite{NJP.15.113022.2013}
\begin{equation}
  \ket{H} = \prod_{e \in E} C_{e} \ket{+}^{\otimes n},
\end{equation}
where $e \in E$ represents an hyperedge.
%
%
%
%
%
As a first example, consider the hypergraph  $(b)$ in the figure. $(\ref{Fig1})$. The quantum state corresponding to this hypergraph is written as
\begin{align}
    \ket{H}=&\frac{1}{\sqrt{8}}(\ket{000}+\ket{001}+\ket{010}+\ket{100} \nonumber\\&
    +\ket{011}+\ket{101}+\ket{110}-\ket{111}).
\end{align} 
Notice that, due to the generalized definition of the $C_z$ gate, only the last part is affected.
Another example is the state corresponding to the 4-uniform hypergraph $(c)$ at figure $(\ref{Fig1})$ since it has four hyperedges of the same cardinality.
The hypergraph state is given by:
\begin{align}
  \label{eq:H17}
  \ket{H_{b}} =
  &
  C_{\{1,2,3\}}C_{\{2,3,4\}}C_{\{1,3,4\}}C_{\{1,2,4\}} \ket{++++}\nonumber \\
  =
  &
    \frac{1}{16}(\ket{0000}+\ket{0001}+\ket{0010}+\ket{0100}+\nonumber&\\&
    +\ket{1000}+\ket{0011}+\ket{0110}+\nonumber&\\&
    +\ket{1100}+\ket{0101}+\ket{1010}+\nonumber&\\&
    +\ket{1001}-\ket{0111}-\ket{1011}-\nonumber&\\&
    -\ket{1101}-\ket{1110}+\ket{1111}).
\end{align}

\begin{figure}[t]
  \centering
  \includegraphics[width=12cm]{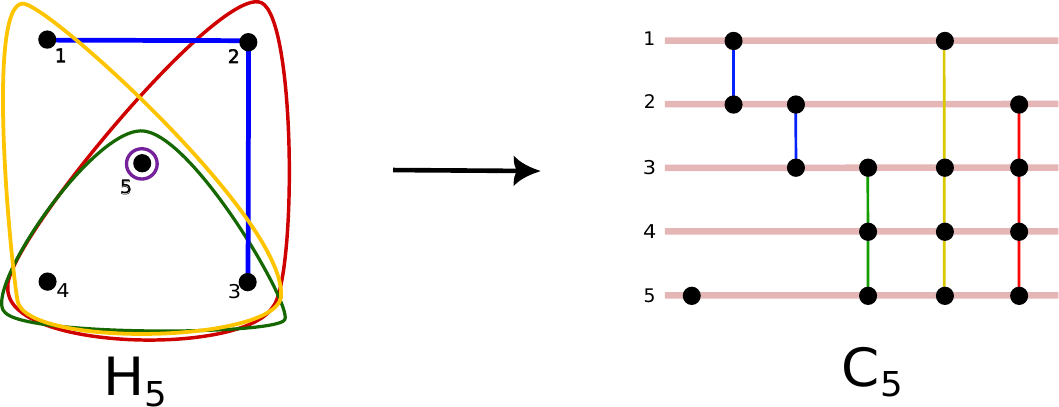}
  \caption{Illustration of the correspondence between a non-uniform hypergraph state $H_5$, on the left, and its associated quantum state, represented in the five-level quantum circuit $C_5$.}
  \label{circuit}
\end{figure}

The correspondence between a hypergraph state and a quantum state associated to a circuit is illustrated in figure (\ref{circuit}), and a novel work proposing a Clifford hierarchy circuit-based scheme for tailoring and learning noise in hypergraph states can be found in details in \cite{lowoverhead}. A key task in quantum information consists of quantifying entanglement for states with many qubits, making use of hypergraph states, which characterize a class of maximally entangled quantum states. 

\begin{figure}[t]
  \centering
  \includegraphics[width=12cm]{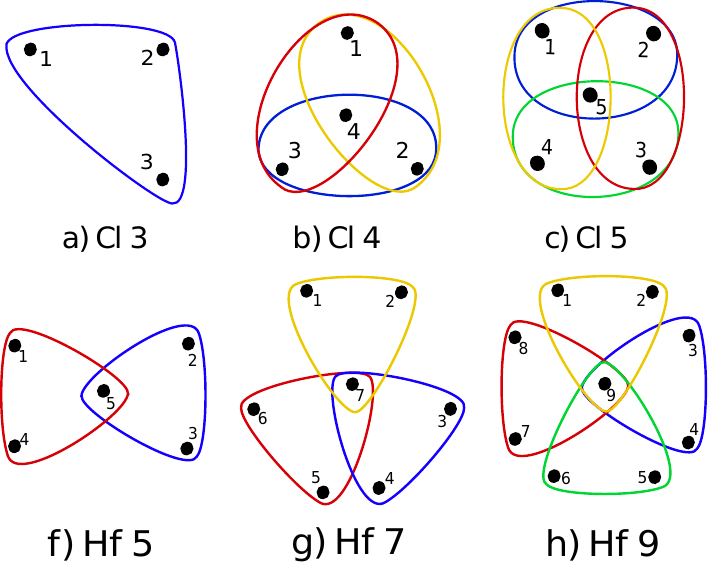}
  \caption{$3$-uniform (a-d) Clover hypergraphs $Cl_n$ and (e-g) the Hyperflower
    hypergraphs $Fl_n$.}
  \label{Fig4}
\end{figure}

Hypergraph states of the Union Jack type present unique properties not seen in graph states, as they offer the possibility of building a universal basis for measurement-based quantum computation using only Pauli measurements \cite{SR.9.1.2019,PhysRevA.99.052304}. 
Entanglement purification protocols for hypergraph states were recently proposed using the CK-DdV\footnote{Carle-Kraus-Dür-de Vicente} protocol, formulated initially for LME states and the only known protocol that works for hypergraph states \cite{PhysRevA.108.062417,PhysRevA.87.012328}. 
Another valuable approach for quantifying entanglement in random pure hypergraph states and their interesting relations to quantum chaos is described in \cite{PRA.106.012410.2022}.
Concerning practical implementation, the application of multiqubit gates is similar to that applied for graph states.  Still, since entangling gates are often probabilistic, the difficulty increases as more qubits are supposed to be entangled at once. Single-qubit
gates usually present fidelities  higher than $99$\%,  two-qubit
entangling gates are around 93\%, and these rates tend to decrease for
more than three-qubit gates
\cite{PRA.83.042314.2011}.
For this purpose, there are practical attempts to implement repeat-until-success methods for more than two qubits using ancilla-mediated multiqubit measurements \cite{finkelstein2024universal,PRL.95.030505.2005}. Different families of hypergraphs may be suitable for measurement-based quantum computation
Particular families of hypergraphs are the Clover hypergraphs, where the k-uniform hyperedges share $k$ vertices and a central vertex shared by all the hyperedges, and Flower hypergraphs, where each hyperedge shares only a common central vertex, as shown in figure (\ref{Fig4}) \cite{andreotti2022}. The clover hypergraph state was successfully implemented experimentally in \cite{vigliar2021error} on silicon-photonic quantum chips, based on previous successful implementations of graph states and indicating that the construction of hypergraph states is also viable \cite{adcock2019programmable,huang2023chip}.

\subsection{Qudit hypergraph states}

High-dimensional entanglement with qudit states, whose state spaces have dimension greater than two, is a challenging yet equally interesting topic. Multiphoton entangled systems for high dimensions have been successfully applied and are enabling future technologies, such as teleporting the quantum information stored in a single photon \cite{erhard2020advances}.

\begin{figure}[t]
  \centering
  \includegraphics[width=12cm]{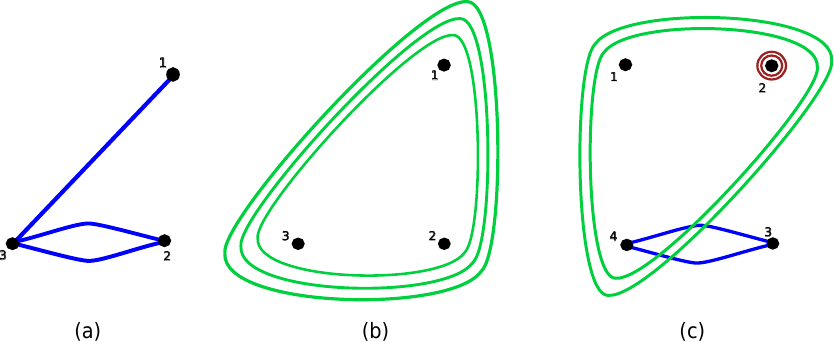}
  \caption{Examples of multi-hypergraph states equivalent to qudit hypergraph states: a) a qudit graph state, b) qudit graph state, c) a non-uniform multi-hypergraph.}
  \label{Fig6}
\end{figure}

A possible generalization of qubit hypergraph states is the use of qudit states.
The d-dimensional Pauli group that stabilizes a hypergraph state for qudits in the computational basis $\{ \ket{0},\ket{1},...,\ket{d-1}\}$ can be defined for the operators $X$ and $Z$ in the form \cite{malpetti2022multipartite}
\begin{equation}
    X=\sum_{q=0}^{d-1} \ketbra{q \oplus 1}{q}
\end{equation}
\noindent and
\begin{equation}
    Z=\sum_{q=0}^{d-1}\omega^q \ketbra{q}
\end{equation}
\noindent that satisfies the conditions $X^d = Z^d = \mathbf{1}$, where $\omega \equiv e^{{i2\pi}/d}$. Thus, we can define the generalized gate $CZ_{\mathcal{I}}$ of $d$ levels as:
\begin{equation}
    CZ_{I}=\sum_{{q_i}_1 = 0}^{d-1}...\sum_{{q_i}_r =0}^{d-1}\omega^{{q_i}_1 ... {q_i}_r}\ketbra{{q_i}_1 ... {q_i}_r}
\end{equation}
\noindent showing that qudit hypergraph states are equivalent to the multihypergraph states, since a qudit would be comparable to a successive application of entangling gates \cite{PhysRevA.95.052340, PhysRevA.97.012323,PhysRevA.109.062407}, as illustrated in the figure (\ref{Fig6}). 
Interestingly, the local gates can be related by a discrete Fourier transform as $X=FZF^{\dagger}$, using the following relation: 
\begin{equation}
   F = d^{-\frac{1}{2}} \sum_{q,q' = 0}^{d-1}\omega^{qq'}\ketbra{q'}{q}   
\end{equation}
\noindent with $Z^d=X^d=I$ and $X^aZ^b=\omega^{-ab}Z^bX^a ~ \forall ~ a,b \in \mathcal{Z}_d$, where $X^aZ^b$ span the d-dimensional Pauli group \cite{steinhoff2025implementation}.

Some research focusing on qudit graph states and applications can be found at \cite{PhysRevA.82.062315, PhysRevA.78.042303}. In the context of many-body interactions, a connection between qudit multi-controlled unitaries and hypergraph states via angular-momentum many-body couplings is also possible \cite{steinhoff2025implementation}. Experimental demonstrations of qudit states are of great interest in the literature for enhancing quantum parallelism in terms of computational capacity, accuracy, and efficiency, in contrast to quantum computing based on qubits \cite{chi2022programmable}.
\subsection{Mixed Hypergraph States}

Although hypergraph states are pure states, in realistic scenarios, we end up with mixed states due to noise or any other undesired interactions. To consider such possibilities, it is possible to construct a species of mixed hypergraph state through a randomization process.
Here, we present the generalization of the method for randomized graph states, based on the Erdos-R\`enyi \cite{PM.6.290.1959,PRA.89.052335.2014} theory of random graphs.
Using a similar procedure, it is emphasized that the hypergraph possibilities involve a larger number of possible combinations, as it may contain many sub-hypergraphs and sub-graphs \cite{PhysRevA.109.012416}.

\begin{figure}[t]
  \centering
  \includegraphics[width=12cm]{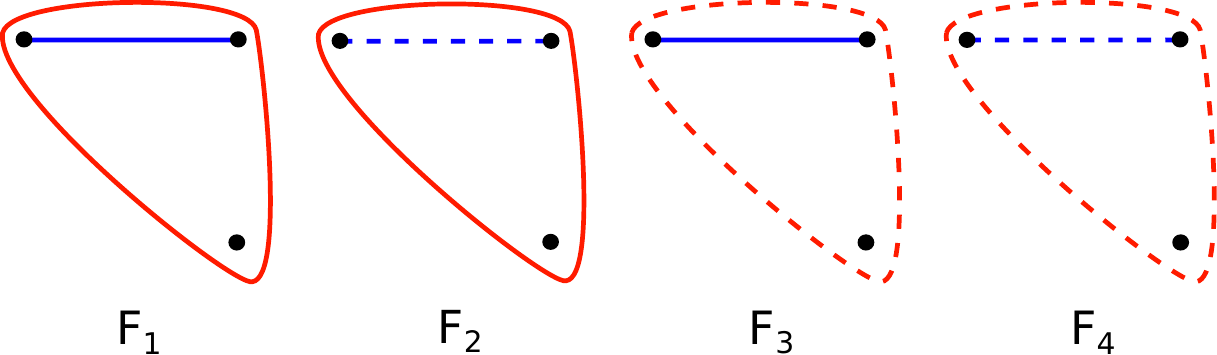}
  \caption{Randomization procedure for the three-qubit hypergraph state and its respective subhypergraphs.}
  \label{random1}
\end{figure}

\begin{figure}[t]
  \centering
  \includegraphics[width=9.5cm]{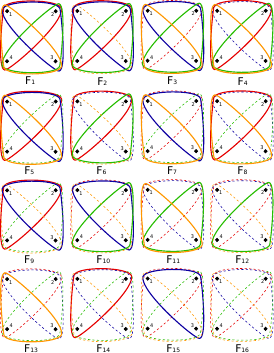}
  \caption{Randomization procedure for the four-qubit hypergraph state and its respective subhypergraphs.}
  \label{random2}
\end{figure}

In hypergraph theory, a sub-hypergraph of a hypergraph $H$ is another hypergraph formed by a subset of vertices and edges of $H$.
Following this decomposition, instead of applying the usual controlled gate, one can define a randomization operator $R_p$ which introduces probabilistic gates  $\Phi^{\{1,2,..., N-qubit\}} _p$ to the separabale state $\ket{+}^{\otimes N}$, where the entangling gate $C_e$ is applied with probability $p$ of success or probability $1-p$ of failure. For example, for the simplest hypergraph, with 3 vertices and one hyperedge, we have:
\begin{align}
    R_p(\ket{H_3})&=\Phi^{[1,2,3]}_p (\ketbra{+++}{+++}) & \\ &=p\ketbra{H_3}{H_3}+(1-p)\ketbra{+++}{+++}\nonumber.
\end{align}
It is natural to conceive the operator as a way to simulate the presence of noise in the system, and as all $\Phi_p$ commute, the application order does not need to be specified. However, the action of this novel operator generates a class of randomized mixed states (RMS), in clear contrast to the usual pure hypergraph states often considered in the literature. The notation for these randomized hypergraph states is usually $\rho^p_H$. Two examples of randomization are presented in the following, for the hypergraph illustrated in figure (\ref{random1}) and for the $3-$uniform hypergraph state with 4 qubits illustrated in figure (\ref{random2}). Notice that the randomization procedure takes the subhypergraphs, which include the subgraphs, in the calculation.
\begin{align}
  R_{P}(\ket{H_{3}})
  & =
  \Phi^{\{1,2,3\}}_{p_3}\circ \Phi^{\{1,2\}}_{p_2} (\ketbra{++++}{++++}) \nonumber \\
  & = p_3 p_2 \ketbra{F_1}{F_1} + p_3 (1-p_2) \ketbra{F_2}{F_2} \nonumber \\ 
  & +p_2 (1-p_3)\ketbra{F_3}{F_3} \nonumber \\
  & +(1-p_3)(1-p_2)\ketbra{F_4}{F_4}
\end{align}
\begin{align}
  R_{P}(\ket{H_{4}})
  = {}
  &
  \Phi^{\{1,2,3\}}_{p_3}\circ \Phi^{\{2,3,4\}}_{p_3} \nonumber \\
  &
    \circ \Phi^{\{3,4,1\}}_{p_3}\circ \Phi^{\{4,1,2\}}_{p_3}
    (\ketbra{+})^{\otimes4}
  \nonumber \\
  = {}
    &
      p_3^{4}\ketbra{F_{1}}{F_{1}}+p_3^{3}(1-p_3)\sum_{k=2}^{5}\ketbra{F_k}{F_k}
      \nonumber \\
    &
      +p_3^{2}(1-p_3)^{2}\sum_{k=6}^{11}\ketbra{F_k}{F_k}
      \nonumber \\
    &
      +p_3(1-p_3)^{3}\sum_{k=12}^{15}\ketbra{F_{k}}{F_{k}}
      \nonumber \\
    &
      +(1-p_3)^{4}\ketbra{F_{16}}{F_{16}},
\end{align}
A more formal definition of a randomized hypergraph state is provided below.
\begin{definition}
  Let $\ket{H}$ be a hypergraph state.
  Its randomization operator is defined as
\begin{equation}
  R_P(\ket{H})= 
  \sum_{F \text{ spans } H}
    \prod_{p_n\in P}
    p_n^{|E_{n,F}|}(1-p_n)^{|E_{n,H} \setminus E_{n,F}|}
    \ketbra{F}{F},
\end{equation}
where $F$ are the spanning subhypergraphs of $H$, $E_{n,H}$ and
$E_{n,F}$ are the sets of hyperedges (of order $\geq 2$) of $H$ and
$F$, respectively, and $P=\{p_k\}_{k=2}^{n}$ is the set of randomization
parameters for hyperedges of order $k$.
The resulting state $\rho_{H}^{P}:= R_P(\ket{H})$ is the randomized
version of $\ket{H}$.
\end{definition}

As stated previously, an operator in the form of a witness of entanglement, as usual in the literature, is not sufficient for calculating the entanglement of mixed and randomized states, since it is necessary to take into account the projector of all the possible combinations of sub-hypergraphs in relation to the original pure state.

Thus, to include the calculus of the witness for the randomized state, we need to compute the overlap $O(\rho_H^p)$, which computes the randomization of all the subhypergraphs, given by 
\begin{align}
    L(\rho_H^p) &:=  Tr[\ket{H}\bra{H}\rho_H^p]\nonumber\\
    & = \sum_{F \hspace{0.1cm} span \hspace{0.1cm} H}p_n^{|E_{F}|}(1-p_n)^{|E_{H} \setminus E_{F}|} \nonumber  Tr[\ket{H}\bra{H}\ket{F}\bra{F}]
\end{align}
\noindent and since the trace is a linear function, the overlap reduces to the
calculation of the scalar product between the original hypergraph state $\ket{H}$ and its subhypergraphs $\ket{F}$.

It is important to highlight the fact that mixed hypergraph states no longer satisfy the stabilizerness condition, i.e., 
\begin{equation}
    h_i R(\ket{H}) \neq R(\ket{H})
\end{equation}
This happens because in the randomization procedure, the free state $\ket{+}^\otimes$ must be taken into consideration. Perhaps such a result might be linked to the fact that two initially pure hypergraph states are non longer equivalent after the randomization procedure, losing the concept of equivalence classes under local operations.

Another interesting but very different approach for random mixed states is presented in \cite{PRXQuantum.2.030347}, where an ensemble of random induced mixed states $\{\rho_A\}$ corresponding to the Hilbert space $\mathcal{H}_A = \mathcal{H}_{A_1} \otimes \mathcal{H}_{A_2}$ is generated by reduced density matrices, which are obtained by
partial tracing random pure states in a composite Hilbert space $\mathcal{H}_A \otimes \mathcal{H}_B$. Such an approach remains applicable to graph and hypergraph states.

\subsection{LME States}

It is essential to notice that graph and hypergraph states are special cases of the locally maximally entangled (LME) states, introduced in \cite{PhysRevA.79.052304}, on which the $2^N$ generalized $C_e$ gate has a more general phase other than $\phi = \pi$:
\begin{equation}
    C_e(\phi)= \mathbf{1}-(1- e^{i \phi})\ketbra{11...1}{11...1},
\end{equation}
\noindent where the phase $\phi$ may represent weighted edges. A recent correspondence of fully connected weighted hypergraphs with non-symmetric GHZ states was established \cite{7zxj-jp34}. By a locally maximally entangled state, we mean a state where each elementary subsystem is maximally entangled with its complement \cite{PhysRevA.90.042308}. LME states are related to weighted hypergraph states, where the weight of each edge corresponds to the phase of its associated operator. 
The condition for the existence of LME states is shown as follows \cite{Bryan2019locallymaximally}.

Let $\mathcal{H}$ be a Hilbert space with elementary subspace dimensions $\vec{d} = (d_1, . . . , d_n)$. There exists a locally maximally entangled state if and only if $R(\vec{d}) \geq 0$, such that
\begin{equation}
    R(\vec{d}) = \prod_i d_i - N(d_1^2,...,d_n^2)
\end{equation}
\noindent where $R(\vec{d})$ is the representation vector, that is, the collection $R_1, ..., R_n$ with dimensions $d_1,...,d_n$ of unitary irreducible representations whose tensor product contains the trivial representation. Famous examples of LME states are the GHZ state and the Bell states.
\section{The stabilizer formalism}
\label{sec4}

\begin{figure}[t]
  \centering
  \includegraphics[width=9cm]{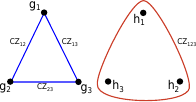}
  \caption{Comparison between the stabilizer operators for a graph and a hypergraph with 3 vertices. Note that the stabilizers for hypergraphs lose the local structure present in graph states. This implies that hypergraph states are no longer in the set of stabilizer states. They are written, respectively for the graph and hypergraph, as $g_1=X_1\otimes Z_2 \otimes Z_3$, $g_2=X_2\otimes Z_1 \otimes Z_3$, and $g_3=X_3\otimes Z_2 \otimes Z_1$, while the stabilizers for the hypergraphs are $h_1= X_1 \otimes CZ_{23}$, $h_2= X_2 \otimes CZ_{13}$ and $h_3= X_1 \otimes CZ_{12}$, emphasizing the nonlocal property of the stabilizers for hypergraph states, since the local Z Pauli is now a generalized controled-Z of dimension proportional to the neighbourhood of vertice $i$. }
  \label{fig:fig8}
\end{figure}

Introduced in \cite{gottesman1997stabilizer} for quantum error correction, the stabilizer formalism provides a setting to demonstrate that operations in quantum computation using the Clifford group are classically efficiently simulable \cite{PhysRevA.73.022334,PhysRevLett.91.147902,PhysRevLett.91.147902}.
As exemplified in the figure (\ref{fig:fig8}), hypergraph states generalize the stabilizers for the graph states in such a way that they acquire a non-local action, making it unlikely to build the stabilizers as merely the tensor product of local operators \cite{fattal2004entanglement}. 

\begin{definition}
    Given a hypergraph $H=(V,E)$ of $N$ qubits and a set of vertices $V={1,2,...,N}$.
A stabilizer is defined on each vertex $i \in V$ as an operator $h_i$ consisting on the tensor product of the Pauli operator $X_i$, applied over the vertex $i$, and the generalized control gate $CZ_{e_j}$ over its adjacent neighbours:
\begin{equation}
    h_i= X_i \bigotimes_{e_j \in A(i)} CZ_{e_j}.
    \label{estabilizer}
\end{equation}
such that $h_i$ stabilize any hypergraph state, that is, satisfy the following property:
\begin{equation}
    h_i \ket{H} = \ket{H}, \qquad \forall \quad i\in V.
\end{equation}
\end{definition}
%

Through the definition of the stabilizing operators, we can define the stabilizer group.

\begin{definition}
    Given a set of $2^N$ operators, the stabilizer group $\mathcal{S}$ is 
\begin{equation}
    \mathcal{S}=\{S_x | S_x = \prod_{i \in V}(h_i)^{(x_i)} \quad \forall \quad x \in \mathrm{Z}\},   
\end{equation}
\noindent which takes all the products between the operators $h_i$. 
\end{definition}

Thus $\mathcal{S}$ is the Abelian group with $2^N$ elements that necessarily satisfy the condition:
\begin{equation}
    \mathrm{S}_x \ket{H}=\ket{H},
\end{equation}
\noindent that is, hypergraph states are not only eigenstates of each stabilizer $h_i$, but also possible combinations of these operators. Now, it is possible to write the hypergraph state basis in the form:
\begin{equation}
    \ketbra{H}{H}=\frac{1}{2^N}\sum_{x \in {\mathcal{Z}_2}^N}S_x = \frac{1}{2} \prod_{i=1}^N(h_i+\mathbf{1}).
\end{equation}
We shall see that such a description is helpful in the formulation of Bell inequalities using the stabilizer formalism of hypergraph states. 

\subsection{Magic or Nonstabilizer states}


Another essential characteristic of hypergraph states is directly related to the fact that they do not belong to the set of stabilizer states, since $CZ_e$
are not Clifford gates as $|e| > 2$. Thus, they are capable of producing the so-called ``magic'', a resource that allows quantum computers to attain an advantage over classical computers \cite{oliviero2022measuring}. The magic or non-stabilizerness of a hypergraph state characterizes the deviation from a given quantum state from the set of stabilizer states, and this is essential to understand the complexity of a state and its usefulness as a resource for quantum computation and error fault tolerance \cite{campbell2017roads,chen2024magic,PhysRevLett.104.030503}. Magic states are a set of ancilla states that enable universal quantum computation beyond Clifford operations \cite{PhysRevA.71.022316}. For systems in continuous variables, the cubic phase state plays the role of the magic
state \cite{PhysRevLett.112.120504}. In practical implementations of measurement-based quantum computation restricted to Pauli measurements, magic can be considered as a resource for completing the Clifford set to a universal set for quantum computation \cite{PRXQuantum.3.020333}.
 
%
Recently, a novel measure, the stabiliser Renyi-Entropy (SRE), was proposed to quantify the magic of a quantum state, although this remains complicated for multiquibit states \cite{PhysRevLett.128.050402}. This measure is quantified in terms of the probability distribution from the projection onto the Pauli operators \cite{chen2024magic} 
\begin{equation}
    P_ {\vec{x},\vec{z}}= \omega(\vec{x},\vec{z})X^{\vec{x}}Z^{\vec{z}}
\end{equation}

\noindent where $\omega$ is a global phase given by $\omega(\vec{x},\vec{z})=\sqrt{-1}^{\sum_{i=1}^nx_i z_i}$ and $\vec{x}$ and $\vec{z}$ are binary vectors with dimension $n$. For pure states $\rho = \ketbra{\Psi}{\Psi}$, we can write:

\begin{align}
    \Tr{\rho^2}=& \sum_{P_i,P_j\in\mathcal{P}_n} 2^{-2n} \Tr{P_i \rho} \Tr{P_j \rho}\Tr{P_i P_j}   \\ 
    = & \sum_{P_i \in P_n} 2^{-n}\Tr{P_i \rho}^2 \\
    =&1
\end{align}
\noindent where the quantum state $\rho$ was decomposed onto the Pauli basis as the Pauli-Liouville representation for a quantum state 
\begin{equation}
    \rho = \sum_{P\in P_n} 2^{-n} \Tr{P \rho}P
\end{equation}
\noindent and $\alpha$ being the usual parameter from the Rényi entropy. Using the expression for $\textbf{m}_\alpha$, defined as the Pauli-Liouville moment,
\begin{equation}
    \textbf{m}_\alpha (\ket{\Psi}) = 2^{-n}\sum_{P \in P_n} (\Tr{P \rho})^{2 \alpha}
\end{equation}
then, the Rényi entropy is therefore given by \cite{chen2024magic}
\begin{align}
    M_{\alpha}(\ket{\Psi})=&\frac{1}{1-\alpha}\log\sum_{P\in\mathcal{P}_n}\left( 2^{-n} \Tr{P\ketbra{\Psi}}^2 \right)^\alpha -n \\
    =&(1-\alpha)^{-1}\log \textbf{m}_{\alpha}(\ket{\Psi}).
\end{align}
%
Another approach for studying magic of hypergraph states, mainly using the measure called min-relative entropy $D_\textbf{min}(\ket{\Psi})$ in order to relate the quantum fidelity to the stabilizers, can be found in \cite{PhysRevLett.131.180401}. Interestingly, random pure states typically reach the maximum magic. A recent experiment observed magic states in top-quark pairs produced in proton-proton collisions, opening a new avenue of convergence between quantum information and particle physics \cite{cms2024observation}.

\subsection{The generalized stabilizer formalism}
An alternative way to write the stabilizers of hypergraph states is to express the generalized controlled gate in terms of local operators and to formulate the nonlocal stabilizer formalism in a way that applies to uniform complete hypergraphs \cite{CHSHhypergraphs}.
To this end, we use the generalized $CZ$-gate in the following form.
\begin{align}
C_{\mathcal{S}} = \mathbb{I}^{\otimes|\mathcal{S}|} - 2\prod_{i\in \mathcal{S}}\ket{1}\bra{1}_{i},
\end{align}
\noindent where $\mathcal{S}$ is a set of qubits. For the stabilizer $h_{i}$, we have $\mathcal{S}\subseteq V\backslash\{i\}$ where $V$ is the the vertex set. Considering $Z\ket{0}=+1\ket{0}$ and $Z\ket{1}=-1\ket{1}$, we have 
\begin{align*}
\ket{0}\bra{0}=\frac{\mathbb{I}+Z}{2},
\qquad
\ket{1}\bra{1}=\frac{\mathbb{I}-Z}{2}.
\end{align*}
Thus we can write $C_\mathcal{S}$ as follows: 
\begin{align}
C_{\mathcal{S}} 
&= 
\mathbb{I}^{\otimes|\mathcal{S}|} - \frac{2}{2^{|\mathcal{S}|}}\prod_{i\in \mathcal{S}}(\mathbb{I}_{i}-Z_{i})
\nonumber\\
&= 
\mathbb{I}^{\otimes|\mathcal{S}|} 
- 
\frac{1}{2^{|\mathcal{S}|-1}} 
\sum_{s\subseteq \mathcal{S}}(-1)^{|s|}\prod_{j\in s}Z_{j}
\nonumber\\
&= 
\mathbb{I}^{\otimes|\mathcal{S}|} 
- 
\frac{1}{2^{|\mathcal{S}|-1}} 
\sum_{s\subseteq \mathcal{S}}(-1)^{|s|}Z_{s},
\end{align}
where $s$ is the subset of $\mathcal{S}$ and $Z_{s}$ is given by
\begin{align}
Z_{s} = 
\begin{cases} 
Z_{\emptyset} = \mathbb{I},  & |s|=0
\\
\prod_{i\in s}Z_{i}
& |s|>0
\end{cases},
\end{align}
taking apart the term associated to the empty set of the summation, it assumes the form: 
\begin{align}
C_{\mathcal{S}} 
&= 
\mathbb{I}^{\otimes|\mathcal{S}|} 
- 
\frac{1}{2^{|\mathcal{S}|-1}} 
\left[ 
\mathbb{I}^{\otimes|\mathcal{S}|} + \sum_{\substack{s\subseteq \mathcal{S} \\ s \neq \emptyset}}(-1)^{|s|}Z_{s}\right]
\nonumber\\
&= 
\left(
1 
- 
\frac{1}{2^{|\mathcal{S}|-1}} 
\right)\mathbb{I}^{\otimes|\mathcal{S}|}
+
\sum_{\substack{s\subseteq \mathcal{S} \\ s \neq \emptyset}}\frac{(-1)^{|s|-1}}{2^{|\mathcal{S}|-1}} Z_{s}.
\end{align}
This form can be expressed more robustly:
\begin{align}
C_{\mathcal{S}} 
&=
\sum_{s\subseteq \mathcal{S}}z_{|s|}(|\mathcal{S}|)Z_{s},
\end{align}
where the constant $z_{|s|}(|\mathcal{S}|)$ is
\begin{align}
z_{|s|}(|\mathcal{S}|)
&= 
\begin{cases} 
&1-\frac{1}{2^{|\mathcal{S}|-1}}, ~\,\quad |s| = 0
\\
&\frac{(-1)^{|s|-1}}{2^{|\mathcal{S}|-1}}, \qquad |s|>0
\end{cases}.
\nonumber\\
&=
\delta_{|s|,0} + \frac{(-1)^{|s| - 1}}{2^{|\mathcal{S}| - 1}}.
\end{align}
Another possible notation in the sum operator is $s\in\mathbb{P}(S)$ with $\mathbb{P}(S)$ being the power set of $\mathcal{S}$ which contains $2^{k-1}$ elements. 

In order to write the stabilizers of a hypergraph state, we must be able to express the products of generalized $CZ$ gates. Then, we have
\begin{align}
\prod_{i=1}^{k} C_{\mathcal{S}_i} 
= 
\sum_{s \subseteq \bigcup_{i=1}^{k} \mathcal{S}_{i}} 
c_{s}
Z_s
\end{align}
\noindent where
\begin{align}
c_{s} = 
\sum_{\substack{s_1\subseteq \mathcal{S}_{1}, \cdots, s_k\subseteq \mathcal{S}_{k} \\ s_1 \bigtriangleup \cdots \bigtriangleup s_k = s}} \prod_{i=1}^k z_{|s_i|}(|\mathcal{S}_{i}|).
\end{align}
As an example, take again the $3$-uniform hypergraph with $N = 4$ vertices. The stabilizer associated with vertex~$1$ is given by
\begin{align}
g_{1} &= X_{1}\prod_{e\in N(1)} CZ_{e} = X_1\, CZ_{23}CZ_{24}CZ_{34}.
\end{align}
And, according to the expansion formula for a two-qubit controlled-$Z$ gate,
\begin{equation*}
CZ_{i,j} = \tfrac{1}{2}\bigl(\mathbb{I} + Z_i + Z_j - Z_i Z_j\bigr),
\end{equation*}
we first obtain
\begin{equation*}
CZ_{23}CZ_{24} = \tfrac{1}{2}\bigl(\mathbb{I} + Z_2 + Z_3Z_4 - Z_2Z_3Z_4\bigr),
\end{equation*}
Then, including the third term,
\begin{equation*}
CZ_{23}CZ_{24}CZ_{34} = \tfrac{1}{2}\bigl(Z_2 + Z_3 + Z_4 - Z_2Z_3Z_4\bigr)
\end{equation*}
hence, the stabilizer generator associated with vertex~$1$ takes the form
\begin{equation}
g_1 = \tfrac{1}{2}X_1\bigl(Z_2 + Z_3 + Z_4 - Z_2Z_3Z_4\bigr),
\label{stabilizer H_3^4}
\end{equation}
\noindent demonstrating the writing of the nonlocal stabilizer in terms of local operators, and the same for the other vertices. This method might be helpful in applications such as quantum error correction and stabilizer codes \cite{balakuntala2017quantum,fu2025error}. 

Using a different approach, an extension of the Pauli stabilizer formalism was made in \cite{webster2022xp}, where, by defining an operator $P$ that performs fractional rotations around the $Z$ axis, it is possible to write the stabilizer generators using the $X$ and $P$ operators, allowing operations on the so called XP codes that cannot be classically simulated. Their relation to hypergraph states lies in the fact that the XP states can be mapped to weighted hypergraph states and vice versa.

\section{Local unitary operations and equivalence classes of entanglement}
\label{sec5}

While entanglement can be treated as a resource for manipulating information, operations that do not increase the entanglement of a system are considered as free operations \cite{RevModPhys.91.025001,PhysRevLett.117.020402}, such as local unitary operations and classical communication \cite{chitambar2014everything}, defined as follows.

\textbf{Local unitary equivalence (LU):} Two $N$-partite states $\ket{\psi}$ and $\ket{\phi}$ are equivalent under local unitary operations $U_n$ if there exists a set of unitary matrices that satisfy the relations:
\begin{equation}
    \bigotimes_{i=1}^n U_i \ket{\psi} = \ket{\phi},
    \label{equivalence}
\end{equation}
If eq. (\ref{equivalence}) is satisfied, the states $\ket{\psi}$ and $\ket{\phi}$ have equivalent entanglement, so can be used for the same protocol, with possible losses though.

By local operations on hypergraph states, one means that an operation is made over a specific partition represented by a vertex $v \in V$, and such operations are subclasses of all the completely positive maps (CPM).  
Therefore, one can say that $LU$ is the strongest form of local equivalence.
For hypergraph states, the local unitary operators $U_n \in  SU(2)$ are the generators of the Pauli group $P=\{\mathbf{1}, \sigma_x, \sigma_y, \sigma_z\}$ and have a central role in the alternative definition of hypergraph states through stabilizers.

\textbf{Local Clifford equivalence (LC):} The local Clifford $C_n$ group maps the set of the $n-$fold Pauli group products into itself \cite{PhysRevA.69.022316,tsimakuridze2017graph} and is given by \cite{descamps2024stabilizer,bittel2025complete}
\begin{equation}
    \textbf{C}_n = \{\mathcal{C} \in U_{2^n} | \mathcal{C} ~\textbf{P}_n ~ \mathcal{C}^\dagger=\mathbf{P}_n\},
\end{equation}
\noindent where $\textbf{P}_n$ is the Pauli group given by
\begin{equation}
    \textbf{P}_n=\{ e^{\frac{i \theta \pi}{2}}\sigma_{j_1}\otimes ... \otimes \sigma_{j_n} | \theta, j_k=0,1,2,3\},
\end{equation}
as will be explicited in the local complementation section.
The Clifford group of n-qubits defined in this way contains $2^{n^2+2n+3}\prod_{j=1}^n(4^j-1)$ elements \cite{calderbank1998quantum}.

Clifford circuits can be used to tailor and detect noise in hypergraph states \cite{park2025efficient}. For the
class of standard graph states, it is known that there exist local unitary equivalent states, but not local Clifford equivalent \cite{DBLP:journals/qic/JiCWY10,NJP.15.113022.2013}
As we shall see in detail, the action of the local Clifford operations on hypergraph states can be described by a set of transformation rules known as local complementation. Previous works on graph states remain open to generalization to hypergraphs \cite{PhysRevA.69.022316}.

\textbf{LOCC equivalence:} a local operation is performed in part of the system, and the correlations are coordinated by a classical way to the other part of the system. The operation is considered deterministic if and only if it can be done with unitary probability; otherwise, it is stochastic  \cite{PhysRevA.75.022318}. Deterministic LOCC are also referred to as protocols or superoperators. In the second case, the purpose is to transform a state $\ket{\psi}$ into another state $\ket{\phi}$ with non-zero probability that the operation fails.

The implementation of an SLOCC over a Hilbert space $H_d$ can be described by an invertible matrix belonging to the general linear group $GL_d$ of dimension $d$. Any LOCC operation can be represented as a separable operation performed over a mixed bipartite state $\rho_{AB}$ defined as \cite{PhysRevLett.78.2275,PhysRevA.92.042329}:
\begin{equation}
\mathcal{L}(\rho_{AB})=\frac{\sum_i A_i \otimes B_i\, \rho_{AB}\, A_i^{\dagger}\otimes B_i^{\dagger}}{Tr[\sum_i A_i \otimes B_i\, \rho_{AB} \, A_i^{\dagger}\otimes B_i^{\dagger}]},
\end{equation}
The set of LOCC is a subset of SLOCC, and since $SLOCC \supseteq LU$, two equivalent states under $LU$ shall also have equivalence under SLOCC. Such operations allow one to understand the equivalence between quantum states. 

\section{Equivalence classes under local operations}
\label{sec6}

One of the primary purposes of discovering states that belong to the same class is to ensure that they are locally equivalent, i.e., connected by a local unitary transformation (a local change of basis), thereby sharing the same entanglement properties. Since local operations cannot increase the global entanglement of a quantum system, two states represented by equivalent hypergraphs have the same degree of entanglement. The GHZ and W states from equations (\ref{ghz}) and (\ref{W}) are famous examples of inequivalent states, because one cannot transform a GHZ state into a $W$ state, while any other entangled state can be transformed into a GHZ or $W$ state \cite{PhysRevA.62.062314}.

For hypergraphs, the $LU$ equivalence and $LC$ equivalence are important notions because they are closely related to the concept of isomorphism. In terms of entanglement, two isomorphic hypergraphs correspond to quantum states that possess the same degree of entanglement, and one can go from one to another only by applying local operations, meaning that they belong to the same class of equivalence. The number of equivalence classes grows exponentially with the number of qubits. For two qubits, there is only one class of equivalence.

\begin{figure*}[t]
  \centering
  \includegraphics[width=14cm]{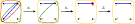}
  \caption{Local unitary equivalence between hypergraph states.}
  \label{fig:fig3}
\end{figure*}

For graph states, it is well known that different graphs lead to locally equivalent states \cite{PhysRevA.80.012102}, and the approach was generalized for hypergraph states \cite{JPA.47.335303.2014,PhysRevA.87.032329}. 
For a hypergraph state with three vertices and one hyperedge, there is only one class by local unitary (LU) equivalence \cite{JPA.47.335303.2014}. Starting from eq. (\ref{estabilizer}), one can write:
\begin{equation}
    X_i \bigotimes_{e_j \in A(i)} C_{e_j}\ket{H}=\ket{H},
\end{equation}
\noindent that is,
\begin{equation}
    X_i \ket{H}=\bigotimes_{e_j \in A(i)} C_{e_j}\ket{H},
\end{equation}
\noindent such that the Pauli operator $X_i$ acts over the qubit $i$ adding all the edges around the adjacency of $i$. The Pauli gate $Z$ is trivial because it only applies the local gate $C_i$ to the target qubit. For the case with three qubits, there is only one $LU$ equivalence class of hypergraph states, represented by the hypergraph (b) in figure (\ref{Fig1}).
For four qubits, however, there are 29 classes of non-equivalent hypergraph states
 \cite{JPA.47.335303.2014}. Figure (\ref{fig:fig3}) shows a practical example of applying the local rules.

Nevertheless, for randomized mixed hypergraph states, the action of a randomization operator onto the state makes these states no longer obey $LU$ and $LC$ equivalence rules, so given two equivalent pure hypergraphs, their randomized versions do not necessarily present the same degree of entanglement, so one must be more suitable for quantum computation. Once they are randomized, those states can not be considered equivalent. In fact, the randomization procedure that leads to the loss of equivalence classes is related to the break of monotonicity of entanglement for any non-uniform hypergraph, so each state must be taken into account when the purpose is to reach specific degrees of entanglement. Notice that the hypergraph states used as examples in (\ref{random2}) present different entanglement after randomization \cite{salem2026randomized}.

 \section{Local complementation on hypergraph states} 
 \label{sec7}

Local complementation is a fundamental operation for graph and hypergraph states, since it allows one to change a hypergraph's topology by performing local operations on its vertices. These operations belong to the Clifford group, as stated before. 
The basic idea of local complementation is given as follows.

Local complementation acts on a hypergraph by picking out a vertex $i$ and inverting the neighborhood $N_i$ of $i$, i.e., vertices in the neighborhood which were connected become disconnected and vice versa. The local complementation acts on the hypergraph state as a local unitary transformation of the Clifford type, and therefore leaves the nonlocal properties invariant.
Unitary operations acting on qubits are represented by quantum gates, which can act on single qubits or multiple qubits. For local complementation on hypergraph states, the main single-qubit gates are the Hadamard gate, given by:
\begin{equation} H = \frac{1}{\sqrt{2}}
\left( \begin{array}{cc}
1 & 1 \\
1 & -1 
\end{array} \right), 
\end{equation} 
\noindent who maps the states from the computational basis $\{\ket{0},\ket{1}\}$ into a superposition of these states, $\frac{1}{\sqrt{2}}(\ket{0}\pm \ket{1})$ and is equivalent to a 90º rotation around the $Y$ in the Bloch sphere followed by a 180º rotation around the $X$ axis, such that $H = X Y^{1/2}$, and the single qubit gates of the Pauli group,
\begin{equation} X = 
\left( \begin{array}{cc}
0 & 1 \\
1 & 0 
\end{array} \right)
Y = \left(\begin{array}{cc}
    0 & -i  \\
    i & 0
\end{array}\right)
Z = \left(\begin{array}{cc}
    1 & 0  \\
    0 & -1
\end{array}\right).
\end{equation} 

The local complementation operators $X$ and $Z$ can be written in the Clifford group as:
\begin{equation}
X^{1/2}=\frac{1}{2} \left( \begin{array}{cc}
\pm i & \mp i \\
\mp i & \pm i
\end{array} \right) = \ketbra{+}{+}\pm i \ketbra{-}{-},
\end{equation}
\label{clifford}
\noindent and
\begin{equation}
Z^{1/2}= \left( \begin{array}{cc}
1 & 0 \\
0 & \pm i
\end{array}\right),
\end{equation}
\noindent where $\ket{\pm} = \frac{1}{\sqrt{2}}=(\ket{0}\pm \ket{1})$ is the eigenstate of the Pauli operator $X$.  The operators of the Clifford group  $C_n = \{ V \in U_{2^n} | V P_n V^{\dagger}=P_n \}$ stabilize (that is, generate eigenstates with eigenvalue $+1$) the operators of the Pauli group  $\in SU(2)$, so the Clifford group is essential for the local complementation operations in graph and hypergraph states. 
Local complementation was recently extended to hypergraphs \cite{gachechiladze2017graphical}. First, define the set of adjacency pairs in relation to a given vertex $a$ as:
\begin{equation}
    A_2(a)=\{\{e_1,e_2\}|e_1\neq e_2, \quad e_1, e_2 \in A(a)\}, 
\end{equation}
\noindent to distinguish from the other pairs. The local complementation in relation to $a$ comes to complement the edges in the multiset $P=\{e_1 \cup e_2 | \{e_1,e_2\} \in \mathcal{A}_2(a) \}$. By complementation, we mean that the hyperedges are either deleted from the hypergraph if present, or added to it if absent. In this way, the complementation is implemented locally over some vertex $a \in V$ of a hypergraph $H=(V,E)$ by the square root of the stabilizing operators $h_a$.
\begin{equation}
    \tau(a)=\sqrt{X_a}^{\pm} \bigotimes_{e \in A(a)} \sqrt{C_e}^{\mp},
\end{equation}
\noindent where $\sqrt{C_e}=\mathbf{1}-(1-(\pm{i}))\ketbra{11...1}{11...1}$ is the diagonal operator that acts over the vertices that belong to the hyperedge $|e|$. Thus, for any hypergraph state associated with a hypergraph $H=(V,E)$, the transformation $\tau(a)$ over the vertex $a \in V$ operates a local complementation between the pairs of hyperedges on the same hypergraph. A generalization of local complementation can be implemented in quantum networks using party-local Clifford transformations for stabilizer states \cite{englbrecht2022transformations}.

\begin{figure}[t]
  \centering
  \includegraphics[width=7cm]{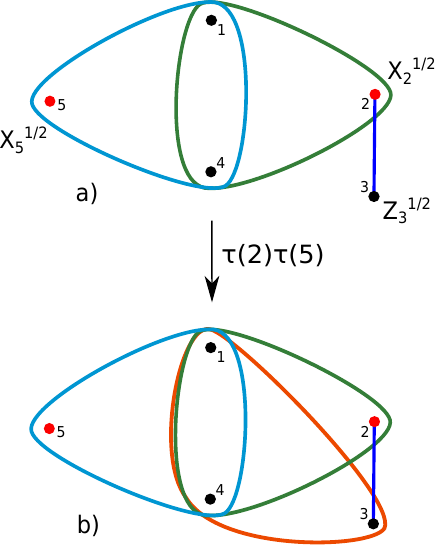}
  \caption{Example of local equivalence of two five-qubit hypergraph states: (a) Two local Clifford operators $\sqrt{X}$ and $\sqrt{Z}$ are applied, responsible for the local complementation on vertices 2 and 5. (b) The hyperedge \{1,3,4\} is obtained as a result of the local complementation on the first vertex. This example shows that, for 5-qubit hypergraph states, local Pauli operators are insufficient to classify unitary local equivalences, because hyperedges form multisets \cite{gachechiladze2017graphical}.}
  \label{fig:fig2}
\end{figure}

\section{Bell inequalities using hypergraph states}
\label{sec8}

One of the main interests in studying hypergraph states is the construction of Bell inequalities and nonlocality arguments, such as the GHZ and Hardy-type arguments \cite{RevModPhys.86.419}.
The approach presented in \cite{PRL.116.070401.2016} allows us to build arguments on non-locality using the stabilizer formalism. Even though stabilizing operators for hypergraph states are nonlocal, they exhibit perfect correlations with respect to certain local measures.

For example, for the hypergraph state with three qubits $\ket{H_3}$, the stabilizer operators are:
\begin{subequations}
\begin{align}
& h_1 = X_1 \otimes C_{23} \\
& h_2 = X_2 \otimes C_{13} \\
& h_3 = X_3 \otimes C_{12},
\end{align}
\end{subequations}
\noindent or, in an explicit form, $h_1$ is written as:
\begin{equation}
    h_1= X_1 \otimes \left( \begin{array}{cccc}
1  & 0 & 0  & 0  \\
0 & 1 & 0 & 0   \\
0 &  0  & 1  & 0 \\
0  & 0  & 0  & -1
\end{array}  \right).
\end{equation}
Since the hypergraph states are permutationally symmetric, it is enough to calculate the correlations of one generator of the stabilizer group. Notice that the elements of the superior diagonal of the operator $C_{23}$ can be considered as projections of the four subspaces that correspond to four results of measurement if we make Pauli measurements $Z$ on the qubits $2$ and $3$, that is, $\ketbra{00}{00}, \ketbra{01}{01}, \ketbra{10}{10}, -\ketbra{11}{11}$, respectivelly.

Therefore, if the action of the first operator over the state of the Pauli basis $X$
generates the result $+$,
\begin{equation}
    h_1 \ket{H_3}=+\ket{H_3}.
    \label{stabilizerr}
\end{equation}
\noindent Thus, the second and third operations cannot generate outputs equal to 1, since this would result in a final projection with a negative value, violating the equation (\ref{stabilizerr}). In terms of probability, we can write in the form:
\begin{equation}
    P(+--|XZZ)=0,
    \label{inadmissivel},
\end{equation}
\noindent Similarly, for the first partition, if it measures in the $X$ basis and yields a $-$ result, the only acceptable results for the partitions measuring in $Z$ are the projectors $-\ketbra{11}{11}$. For the state $\ket{H_3}$ the probability $P(---|XZZ)=1$ is a possible result. In addition to (\ref{inadmissivel}), another three possibilities violate the equation (\ref{stabilizerr}) and can not occur: 
\begin{equation}
\begin{aligned}
    &P(-++|XZZ)+P(--+|XZZ)&\\&+P(-+-|XZZ)=0,
\end{aligned}
\end{equation}
\noindent completing the forbidden probabilities for the first operator, $h_1$. The other probabilites for $h_2$ and $h_3$, as demonstrated by   \cite{PRL.116.070401.2016}, are the following:
\begin{subequations}
\begin{align}
    & P(-++|XZZ)+ P(--+|ZZX)=0 \\\nonumber
    \\\nonumber
    & P(+-+|ZXZ)+ P(--+|ZXZ)+ \nonumber \\
    &  + P(+--|ZXZ)=0 
    \\\nonumber\\&\nonumber
     P(++-|ZZX)+ P(-+-|ZZX) + \nonumber \\& + P(+--|ZZX)=0,
\end{align}
\end{subequations}
\noindent summing up to twelve perfect correlations for the state $\ket{H_3}$. Similarly, the procedures can be generalized for a hypergraph state with $N$ qubits and a $N$-hyperedge of unique cardinality. Then, it is possible to write the Bell operator as Mermin-like operators in an expression which is permutation symmetric, as done in \cite{PRL.116.070401.2016}, 
\begin{align}
    B_N &= \frac{1}{2}((P+iZ)^{\otimes N} + (P - iZ)^{\otimes N})  \\ & 
    = \sum_{m=0, \text{even}}^N i^m Z_1...Z_m P_{m+1}...P_N + \text{perm.} 
\label{mermin}
\end{align}

\noindent for $P \in \{X,Y\}$. This important result consists in the fact that the quantum value of $B_N$ on the state $\ket{H_N^3}$ is equal to $2^{N-2}$, while the local realistic model is bounded by a factor $2^{N/2}$. The same result was obtained using local Pauli symmetries, since symmetric hypergraph states can be associated with generalized GHZ states for $n$ qubits, the Bell operator can be written as \cite{Noller2023,JPA.50.245303.2017}

Taking the Mermin-like Bell operator (\ref{mermin}), which is permutation symmetric, and applying unitary operators $\sqrt{P_{-}}^\otimes$ or $\sqrt{Z_{-}}^\otimes$, it turns out that the Bell operator is now:
\begin{align}
    \Tilde{B}_N &:= \sqrt{P_{-}}^\otimes B_N^P \sqrt{P_{-}}^\otimes \\&
    = \frac{1}{2}((\textbf{1} +Z)^{\otimes N} + (\textbf{1} - Z)^{\otimes N}) \nonumber&\\&
    = \sum_{m=0, \text{even}}^N \textbf{1}_1 ... \textbf{1}_m Z_{m+1} ... Z_N + \text{perm.}\nonumber
\end{align}
\noindent exploring the possibilities of symmetric hypergraphs as GHZ states.

Despite using a different approach, hypergraph inequalities are useful for studying quantum contextuality \cite{e21111107}.

\subsection{Bell inequalities and entanglement witnesses}
The entanglement witness $W$ is a Hermitian operator designed for the detection of entanglement \cite{RMP.81.865.2009}. and it can be separated into two positive operators $P_M + Q_M^cM$ and with $Tr(W) = 1$, satisfying the condition $P_M > Q_M$ \cite{chruscinski2014entanglement,Eisert_2007}. 

The connection between entanglement witnesses and Bell inequalities was first made in \cite{PLA.271.319.2000}. Since they cannot detect all the entangled states, Bell inequalities can be seen as nonoptimal entanglement witnesses. In fact, a CHSH witness that is positive on all LHV states can be constructed as follows:
\begin{equation}
    W_{CHSH}= 2 \mathbb{I}-\mathcal{B}_{CHSH}.
\end{equation}
\noindent so that a Bell inequality is a double witness, which is quite appealing because it detects not only entanglement but also nonlocality. For multipartite systems, Bell inequalities can even detect bound entanglement. However, due to the large ``degree of freedom'' inherent to a Bell inequality, establishing the exact relations to entanglement witnesses is quite complex \cite{PhysRevA.72.012321}.

A possible entanglement witness detecting the entanglement in a general hypergraph state is of the type
\begin{equation}
    \mathcal{W}=\alpha_s \mathbf{1}-\ketbra{H}{H},
\end{equation}
\noindent where $\alpha_s$ is the maximal squared overlap between $\ket{H_i}$ and any biseparable
pure state $\ket{\phi_i} = \ket{\alpha_J} \ket{\beta_{\bar{J}}}$ where $J|\bar{J}$ is a bipartition
of the $N$ qubits. This overlap $\alpha_S$ can be computed directly as the maximum of the eigenvalues of the reduced states. 

For example, for the three-qubit pure hypergraph state $\ket{H_3}$,
\begin{equation}
\begin{aligned}
    \ket{H_3} = & \frac{1}{\sqrt{8}}
    (\ket{000}+\ket{001}+\ket{010}+\ket{100}   \\ & +\ket{011}+\ket{101}+\ket{110}-\ket{111}),
\end{aligned}
\end{equation}
\noindent and by applying a Hadamard gate on the third qubit, we obtain:
\begin{equation}
    \ket{H_3}= \frac{1}{2}(\ket{000}+\ket{010}+\ket{100}+\ket{100})
\end{equation}
so we have for the three-qubit state $\alpha_S = 3/4$. In general, one can show that for N-qubit hypergraph states consisting of only a single
$N-$edge, one has that $\alpha_S=1-2^{(1-N)}$.

For a mixed random hypergraph state, it is necessary to compute the randomization of the superposition $O(\rho^p_H)$ that projects all the respective sub-hypergraphs, given by $Tr[\ketbra{H}{H}\rho^p_H]$. Thus, a general superposition for any hypergraph is written as:
\begin{align}
  O(\rho^p_H) = {}
  & \Tr\left(\ketbra{H}{H}\rho^p_H\right) \nonumber \\
  = {}
  & \sum_{F\text{ spans }H} p^{|E_F|}(1-p)^{|E_H/E_F|}
    & \\ & Tr[\ketbra{H}{H}\ketbra{F}{F}],
\end{align}
\noindent that is, it is necessary to calculate the inner product of the hypergraph state and its sub-states due to the randomization. Such procedure is not trivial and demands considerable computational power due to the large number of possible combinations \cite{salem2026randomized}.

\subsection{Hypergraph States in Quantum Computation}

The application of hypergraph states in quantum computing is related to the ability to verify them using local measurements. Recently, a simple method for verifying hypergraph states using only two Pauli measurements per subsystem was proposed \cite{PhysRevA.96.062321,PRApp.12.054047.2019, SR.9.1.2019,tao2022verification}. By verification, we mean a sequence of single-qubit Pauli measurements on a given state to check whether it is close to an ideal state or not \cite{PhysRevX.8.021060}.
Recently, it was demonstrated that there is a correspondence between hypergraph states and Real Equally Weighted (REW) states, which are used in the implementation of quantum algorithms such as Grover's and Deutsch-Jozsa's algorithms. The search problems that can be simulated efficiently by the Grover algorithm must present the following properties:

\vspace{0.3cm}
1) The system does not possess a searchable structure in the possible answers collection.

2)  The number of possible answers to be verified corresponds to the number of inputs in the algorithm.

3) There is a Boolean function of the type $f: X \rightarrow{B}$ that solves each input and determines if this is a correct response.

\vspace{0.3cm}

The action protocols that define the utilization of algorithms like this can be prepared in the following initial state:
\begin{equation}
    \ket{\psi_0}\equiv \frac{1}{\sqrt{2^n}} \sum_{x=0}^{{2^n}-1}\ket{x},
\end{equation}
\noindent corresponding to a superposition of REW states of all the possible $2^n$ states $\ket{x}$ in the computational basis. Applying an unitary transformation $U_f$ on the target qubit,
\begin{equation}
    U_f \ket{x}\ket{y}=\ket{x}\ket{f(x)\oplus y},
    \label{boolean}
\end{equation}
The following state is generated: 
\begin{equation}
    \ket{\psi_f}\equiv\frac{1}{\sqrt{2^n}} \sum_{x=0}^{{2^n}-1}(-1)^{f(x)}\ket{x},
    \label{rew}
\end{equation}
Where $f(x)$ is a Boolean function, this way, it is possible to define hypergraph states in terms of a Boolean function in the form:
\begin{equation}
    \ket{H}=\frac{1}{2^{N/2}}\sum_{\mathbf{x}\in \{0,1\}^N}(-1)^{f(x)}\ket{\mathbf{x}},
    \label{rew}
\end{equation}
\noindent establishing a biunivocal correspondence between the class of hypergraph states and REW states. In fact, we can observe that the hypergraph with three vertices affected by a unique hyperedge can be expressed by the function $f(x) = x_1 x_2 x_3$. At the same time, graph states can correspond to functions with two variables at most, given that graphs have a maximal cardinality of two. 
An example with three qubits for a REW state $\ket{f}$ is given by. This way, the hypergraph corresponding to a REW state can be obtained by applying a sequence of logic gates. In the cited example, the applications of transforms $Z_1, C^2, Z_{23}$ and $C^3 Z_{123}$ takes to the initial state $\ket{+}^3$. 
For the initial hypergraph state with three qubits, we can apply a Hadamard gate, and the state becomes
\begin{equation}
\begin{aligned}
    \ket{f}=&\frac{1}{\sqrt{8}}(\ket{000} + \ket{001} +\ket{010} - \ket{011}\\ & - \ket{100} - \ket{101} - \ket{110} - \ket{111}),
    \label{rewequation}
\end{aligned}
\end{equation}
\noindent and after the partial trace for the reduced matrices of $\ketbra{H}$, we have eigenvalues $\lambda_1 = \frac{3}{4}$ and $\lambda_2 = \frac{1}{4}$, whose geometric measure of entanglement is $E_2 \geq \frac{1}{4}$. Since any connected graph state has a geometric measure of entanglement $E_2 \geq \frac{1}{2}$, this class of states cannot contain all the REW states, in contrast with hypergraph states. While there is a one-to-one correspondence between qubit hypergraph states and REW states, qudit hypergraph states are a subset of the so-called generalized REW states \cite{PhysRevA.109.062407}. An innovative application is presented in \cite{PhysRevResearch.2.013322}, where a quantum, decentralized blockchain protocol is proposed that uses weighted hypergraph states to record information about classical blocks, since decentralization is one of the features that make a blockchain a distributed ledger with many security and privacy advantages over traditional databases \cite{yang2022decentralization}. The objective of the quantum protocol proposed in \cite{PhysRevResearch.2.013322} is to replace the classical ledger-based network and cryptographic hash with quantum entanglement in a weighted (i.e., hyperedges phase) hypergraph state, where the peers use a quantum channel 
to build the chain according to a mutually agreed-upon consensus. For communication over the quantum channel, the peers can use any quantum key distribution protocol \cite{kiktenko2018quantum,kumar2025brief}. For the construction of the blocks in the chain, the weighted
hypergraph of $n$ qubits represents the $n$ blocks, where each qubit is encoded with the information of a classical block. Once the classical
information of a block is encoded in a qubit, the peer sends a copy of the state to its partners in the network.
\begin{figure}[t]
  \centering
  \includegraphics[width=5cm]{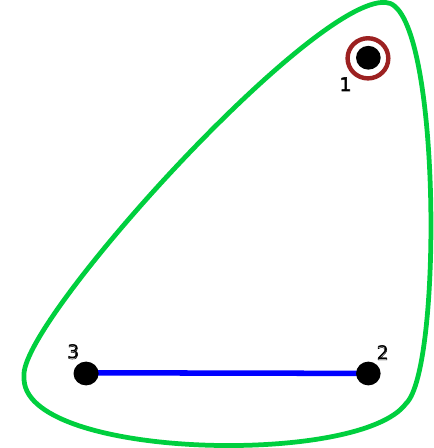}
  \caption{Hypergraph corresponding to the REW state described by equation (\ref{rewequation}).}
  \label{Fig5}
\end{figure}
Quantum error correction is another practical application of multiqubit states, as previously done for graph states \cite{PhysRevA.78.042303,PhysRevA.105.042418,PhysRevA.65.012308,roffe2019quantum}, and recently generalized for hypergraph states \cite{JPA.51.125302.2018,balakuntala2017quantum}, what can help in the development of implementing fault-tolerant quantum computation, i.e., quantum computers that can operate correctly even in the presence of errors, a technology of immense interest in the present era, since the presence of errors mainly due to noise is one of the biggest challenges for quantum computation \cite{shor1996fault,preskill2018quantum}. Indeed, the hypergraph state device successfully implemented in the experiment by Huang et al. \cite{huang2024demonstration} was a noisy intermediate-scale quantum device.

\section{Hypergraph states in continuous variables}
\label{sec9}

Only recently has the scenario for hypergraph states been extended to continuous-variable systems \cite{PRA.100.062301.2019}, and the literature on graph states within this framework is also not particularly abundant, leaving room for new projects. First, it is helpful to introduce some fundamental notions of the operations for these systems, including graph states.
In contrast to quantum information based on discrete variables, quantum states in continuous variables are described using a Hilbert space of infinite dimension, and one can define $N$ bosonic observables with a continuous eigenvalue spectrum associated to the operator $H^{\otimes^N}=\bigotimes_{k=1}^N H_k$.
These operators can be manipulated using continuous-variable algebra when written in the vectorial form as $\hat{b}:=(\hat{a}_1, \hat{a}^\dagger_1, ..., \hat{a}_N, \hat{a}^\dagger_N)$, and must satisfy the canonical commutation rules \cite{PhysRevLett.89.097901}:
\begin{equation}
    \left[\hat{b}_i,\hat{b}_j\right]=\Omega_{ij}, \qquad (i,j=1,2,...,2N)
    \label{bosonica}
\end{equation}
\noindent where $\hat{b}_i,\hat{b}_j$ are the bosonic operators $\Omega_{ij}$ representing the elements of the matrix $2N \times 2N$:  
\begin{align}
\Omega:=&\bigoplus_{k=1}^N \omega, ~~~~\omega := \left( \begin{array}{cc}
0 & 1  \\
-1 & 0    
\end{array}\right) \nonumber \\= &\left( \begin{array}{cccc}
\omega & & & \\
 &  ... &   & \\
 & &  & \omega    
\end{array}\right),
\end{align}
\noindent known as the sympletic form. This way, any density operator of $N$-modes can be defined over a space of dimension $2N$.


A pure state is Gaussian if and only if its Wigner function is non-negative \cite{PhysRevA.79.062302, HUDSON1974249}. 
Gaussian states are bosonic states that can be described in the complete form by the characterization of the first two \textit{momenta}, including the covariance matrix, that is, $\hat{\rho}=\hat{\rho}(\Bar{x}, \mathbf{V}).$ A Gaussian state is defined as the state whose characteristic function is Gaussian \cite{WANG20071,sansavini2019continuous}: 
%
%
\begin{equation}
\begin{aligned}
    W(x)=&\frac{1}{(2 \pi)^N\sqrt{\det V}} \times \\&
    \exp{-\frac{1}{2}(\mathbf{x}-\mathbf{\Bar{x}})^T\mathbf{V}^{-1}(\mathbf{x}-\mathbf{\Bar{x}})}.
\end{aligned}
\end{equation}

It is worthy to note that the decomposition of a Gaussian state into a thermal state maximizes the Von Neumann entropy:
\begin{equation}
    S:=-Tr(\hat{\rho} \ln{\hat{\rho}}),
\end{equation}
\noindent for a fixed energy $Tr(\hat{\rho} \hat{a}^{\dagger}\hat{a})=\Bar{n}$, with $\Bar{n}\geq 0$ being the average of the number of photons in the bosonic mode \cite{demarie2018pedagogical}.
%
A Gaussian operation transforms Gaussian states into Gaussian states, producing a convex set.
When compared to quantum states in discrete variables, Gaussian states are more easily manipulated experimentally, and considerable progress has been made in manipulating non-Gaussian states \cite{ra2020non,stiesdal2021controlled}. 
In the transposition to graph states as Gaussian states, instead of using the Hadamard basis, $\ket{\pm}$, we take the state $\ket{0}_p$, and measurements on the Pauli basis are replaced by $\hat{p}$, while $\hat{Z}$ is replaced by $\hat{q}$. A graph state $\ket{\phi}$ with adjacency matrix $\mathcal{A}_{ij}$ can be studied as a Gaussian cluster state, and written as \cite{sansavini2019continuous}
\begin{equation}
    \ket{\phi}= \prod_{1\leq i<j <N} \exp{i \mathcal{A}_{ij}q_i \otimes q_j} (t_{e_{i,j}}) \ket{0}^{\otimes N}_p,
\end{equation}
\noindent where the initially uncorrelated set of squeezed modes are then associated with a graph state by using the operator $\exp{i \hat{q_j}\hat{q_k}}$ and $(t_{e_{i,j}})$ is a weight associated to each gate. For more than two vertices, it is equivalent to a Toffoli gate $e^{i \hat{q}_{v_1}...\hat{q}_{v_n}}$ \cite{PhysRevResearch.6.023332}.

For Gaussian cluster states, the generalized coordinates vector, $(\hat{q},\hat{p})$, here defined as $\hat{X}:=\{\hat{q}_1,...,\hat{q}_N;\hat{p}_1,...,\hat{p}_N\}$, satisfies the condition $<\hat{X}>=0$. These can be identified as the position and momentum operators \cite{PhysRevA.111.022413}. 
Analogous to the case of qubit states, Gaussian graph states can be described by their stabilizers as a set of operators written as
\begin{equation}
    g_i(s)=\hat{X}_i(s) \prod_{j\in N(i)} \hat{Z}_j (s),
    \label{stabilize}
\end{equation}
\noindent where $s$ is the squeezing parameter, $X(s) = e^{ips} $ and $Z(s)=e^{-isq}$. 



The stabilizer group is defined by a Lie algebra that constitutes the nullifier space of operators $\hat{H}$ that satisfy the condition
\begin{equation}
    \hat{H} \ket{\phi}=0,
\end{equation}
Every operator $\hat{H}$ that fulfils this condition is hermitian and therefore, an observable. A nulifier $H_i$ acting on the vertex $i$ can be formally defined as \cite{ravikumar2025nonclassical}:
\begin{equation}
    \hat{H}_i=\hat{p}_i-\sum_{j\in N(i)} \hat{q}_j,
\end{equation}
for $i=1,...,n$. Any linear superposition $\hat{H}=\sum_i c_i \hat{H}_i$ satisfies $H\ket{\phi}=0$. Every nulifier also must satisfy the commutation relation $[\hat{H}_i,\hat{H}_j]=0$ for any $(i,j)$, in such a way that any  $H$ in the nulifier space $({H\ket{g_{\leq n}}=0})$ can be written as a linear combination of the nulifiers $H=\sum_j c_j H^{\leq n}_j$. Every nullifier is hermitian and also an observable.

The Wigner function $\mathcal{W}(q,p)$ of a graph state $\ket{\phi}$ with $n$ $\textit{qumodes}$ can be written as \cite{PhysRevA.79.062318}:
\begin{equation}
    \mathcal{W}(q,p)=\prod_{i=1}^n \varepsilon(q_i) \delta(H_i),
\end{equation}
where $\varepsilon(x)$ is the infinite uniform distribution and $H_i$ are the standard nullifiers of $\ket{\phi}$. In the ideal scenario, graph states are highly singular, so the use of $\delta$ and $\varepsilon$, which are the limits of the Gaussian function whose variance goes from zero to infinity \cite{PRA.101.033816.2020}.
%
%
%
In terms of the covariance matrix $\textbf{V}$ of a 2 $m$-dimensional graph state described by a adjacency matrix  $\mathcal{A}$ \cite{PhysRevLett.121.220501}, we have
\begin{equation}
    V_0 \xrightarrow{} V=G^T V_0 G \quad | \quad G=\left( \begin{array}{cc}
        \mathbf{1} & \mathcal{A} \\
         0 & \mathbf{1}
    \end{array}\right)
\end{equation}
\noindent where $V_0= diag(s_1,...,s_m;s_1^{-1},...,s_m^{-1})$.

%

However, graph states produce Gaussian statistics in the quadrature measures of the electromagnetic field, thus such statistics can be simulated in an efficient way using classical computation, so the Gaussian operations over Gaussian states can not generate a universal basis for quantum computation, and neither can they violate a Bell inequality, since some form of non-Gaussianity is required, such as supplement the homodyne detection with some non-Gaussian measurement  \cite{PhysRevLett.88.097904, PhysRevLett.109.230503,RevModPhys.77.513}. For this reason, in the context of resource theories, Gaussian states and Gaussian operations are not considered resource states or operations \cite{PhysRevA.98.022335, PhysRevA.97.062337}. 
That being said, for achieving universal quantum computation, it is possible to introduce non-Gaussian operations such as the cubic phase or photon addition (subtraction) on the state \cite{PRXQuantum.2.010327,walschaers2023emergent}, the first being of complex experimental implementation, turning the second option usually more desired \cite{PhysRevA.84.053802, PhysRevLett.98.030502, PhysRevX.7.031012}. Theoretically, the photon addition (subtraction) is represented by the action of annihilation and creation operators using non-linear optics \cite{PhysRevX.7.031012, PhysRevA.89.063808, PhysRevA.75.052106}. 
Graph states can be constructed in continuous-variable complex networks via passive optics, and a generalization to hypergraph states remains open \cite{fainsin2025entanglement}.

%

%

In the continuous-variable scenario, the main difference for qumode graph states is that qumode hypergraph states are non-Gaussian, satisfying the Lloyd-Braunstein criteria even when restricted to a Gaussian measurement strategy \cite{PRA.100.062301.2019,PhysRevLett.82.1784}. For a hypergraph state in continuous variables, the vertices represent $n$ quantum modes and the weighted hypergraphs represent the generalized controlled-Z gates, as defined in the following.
\begin{equation}
   \ket{H}=\prod_{k=1}^n \prod_{e_k \in E} C_{e_k}(t_{e_k})\ket{0}^{\otimes n}_p,
    \label{null}
\end{equation}
\noindent where $\ket{0}_p$ is the zero-momentum eigenstate for $n$ modes in the limit of infinite squeezing, $C_{i_1,...,i_k}=e^{i q_{i_1},...,q_{i_k}}$, that is, the weighted gate written in terms of exponential, as mentioned for the particular case of graph states. Note that, in the continuous variables approach, we make use of weighted hyperedges, where $t_{e_k}$ are elements of the set of real valued weights $T={\{t_{e_k}\}}^{|E|}_{k=1}$ associated to each hyperedge, where $|E|$ is the cardinality of $E$. Similarly to what occurs in discrete variables, the subindexes indicate over which $\textit{qumodes}$ the operators act. As an example, take the last hypergraph in figure (\ref{fig:fig3}) with four vertices and a hyperedge with cardinality $4$ and a usual edge between the vertices $1$ and $2$. This hypergraph state is given by $e^{i q_1 q_2 q_3 q_4}$ $e^{i q_1 q_2}\ket{0}^{\otimes 4}_p$. 
%
%
The manipulation of non-Gaussian states using a hypergraph framework is analyzed in \cite{vandre2025graphical} and this is important because it can be used as resources \cite{PRXQuantum.2.030204}.



Similarly, hypergraph states in continuous variables can also be described using the stabilizer formalism,
\begin{equation}
    g_i^{(k)}(s)=X_i (s) \bigotimes_{e_{k-1} \in \mathcal{N}(i)} C_{e_{k-1}}(s)(t_{e_k})
\end{equation}
so if $X$ is a stabilizer for a given state $\ket{\psi}$, $UXU^\dagger$ is a stabilizer for $U\ket{\psi}$.

The nullifers for a hypergraph state generalize the ones for graphs:
\begin{equation}
    H_i^{(k)}= p_i -\sum_{e_{k-1}\in \mathcal{N}(i)} q_{i_1}... q_{i_{k-1}}.
\end{equation}

We should note the fact that any Gaussian measurement on a qumode, i.e., a vertex of a hypergraph, produces the effect of reducing in one factor the order of the hyperedges connected to the given vertex.


\subsection{Gaussian operations on hypergraph states}

Since hypergraph states are non-Gaussian states, Gaussian measurements applied on them might be suitable for providing universal quantum computation. 

The displacement operators are given by the same $Z$ and $X$, that when applied on a hypergraph state, generate adittional correlations \cite{vandre2025graphical}

The connection between hypergraph states and a reservoir, which is necessary in this case for a thermal hypergraph state, is a viable implementation in a dynamical system with decoherence. 

For this purpose, we utilize the Hamiltonian $\textbf{H}$,
\begin{equation}
	\textbf{H}= - \sum_{i=1}^n g_i,
\end{equation}
\noindent where $g_i$ are the stabilizers and whose thermal state is given by
\begin{equation}
	\rho_T \equiv \frac{e^{-\beta \textbf{H}}}{Z},
\end{equation}
\noindent where  $Z \equiv Tr[e^{-\beta \textbf{H}}]$ is the partition function that describes the thermal state at a given temperature $T$. If $T=0$, we have the pure state $\ketbra{H}$.

In the case $\mathcal{H}=-\sum_{i=1}^n g_i$, we have  

\begin{equation}
    \rho_T =  \left( \prod_{i=1}^n \Omega_i^{P_\beta}  \right)  (\ketbra{H}{H})
\end{equation}
\noindent where $\Omega$ is a superoperator $\Omega (.)  = (1-p)(.) + p Z_i (.) Z_i $ and $p_\beta \equiv e^{-2\beta}/(1+e^{-2\beta})$. \cite{PhysRevA.106.012405}.

%
\begin{table}[h]\caption{Comparison between the main differences of discrete and continuous variables approaches.}
\centering
\label{table}
\begin{tabular}{lll}
 & \textbf{Continuous variables} &   \textbf{Discrete variables} \\
\\
\textbf{Element} & Qumodes  &  Qubits \\
\\
\textbf{Operators} & $\{\hat{x},\hat{p}\}$; $\{\hat{a},\hat{a}^{\dagger}\}$  & $\{\hat{\sigma}_x$,$\hat{\sigma}_y$,$\hat{\sigma}_z\}$ \\
\\
\textbf{States}  & $\ket{\alpha}$; $\ket{z}$; $\ket{n}$ & Pauli group\\
\\
\textbf{Gates} & $R_i(\phi)$, $D_i(\alpha)$, $S_i(z)$, $BS_{i,j}$ & T-gate, Hadamard, CNOT   \\
\\
\textbf{Measurements} & Homodyne $\hat{x}_\phi$; heterodyne $Q(\alpha)$  & Pauli measurements.  
\end{tabular}
\end{table}
\label{table}
%

\subsection{Overview}

The different approaches presented in this review highlights the fact that hyperrgaph states can be useful both in discrete and continuous description. 
Although the information processed in a quantum computer is usually made with qubits, there is more room for encoding quantum information into optical states, since the bosonic Fock space is inﬁnite dimensional, but creating a Fock state with many photons is a difficult proccess, as well as counting large photon numbers.
Regarding the technical implementations of the two approaches, a key difference lies in the measurements performed on each system. In discrete quantum information, measurements in the Pauli basis are performed, for example, in Measurement-Based Quantum Computation. For Gaussian quantum information, there are typically two types of detection: homodyne and heterodyne. The first method involves extracting the received information by modulating the phase or frequency of an oscillating signal with a pattern identical to the first, if it carries no information. The second detection implies the use of two different frequencies, which justifies its name. An original proposal for a mixture between both approaches in a "hybrid" quantum information can be found in \cite{andersen2015hybrid,van2011optical}, and is exposed in the table (\ref{table}). Such a presentation might be suitable for students who might want to work with graph and hypergraph states in both panoramas.

\section{Concluding remarks and open works}

In this review, we provided a concise tutorial on the theory of hypergraph states and their applications. The emerging literature demonstrates growing interest in multiqubit entangled states such as hypergraph states, and their applications in quantum computation. Despite the difficulty of dealing with multiparticle systems, the usefulness of hypergraph states as resources for quantum protocols is significant, not only because they immediately generalize graph states, but also because they exhibit unique properties that cannot be observed in graph states. The purpose of this concise review is to update the interested reader in all the main works on hypergraph states in the last decade, and to provide useful insights for those looking for open investigation in the topic.

Regarding open work on hypergraph states, many of them concern generalization from graph states.  Considering mixed hypergraph states, if they can yield more realistic scenarios by simulating noise, practical implementations are welcome and the extension to magic in mixed states is still necessary. In qudit states, more approaches are necessary, such as using qudit hypergraph states in noisy channels \cite{dutta2023qudit}, and definitions of qudit hypergraph states for continuous variables as well.
Indeed, hypergraph states with continuous variables enable the extension of many hypergraph applications in discrete variables to the realm of Gaussian quantum information \cite{takeuchi2019resource}.

Another possible research topic, both in the context of discrete quantum information, which has been the main focus of this study so far, and in continuous variables, is related to foundational studies such as Bell inequalities, self-testing, and GHZ-like arguments for nonlocality \cite{PhysRevLett.124.020402, PhysRevLett.95.120405,panwar2023elegant,mckague2011self,PhysRevA.77.032108,PhysRevLett.110.100403}, as well as applications on quantum computing, as transcription of continuous variable states into hypergraph states \cite{PhysRevA.78.052307,RevModPhys.84.621,Adesso2014} state fission \cite{PhysRevA.111.052624}, state preparation \cite{cam2025universal} and photonic quantum computing (PQC) \cite{li2025reinforcement}. 

Recent interest in quantum networks in the literature presents an opportunity to construct continuous-variable networks using hypergraph states, as was recently done for graph states, where entanglement is built using squeezed states and linear optics, providing a valuable framework for implementing many protocols in quantum information, taking squeezed qummodes on a network \cite{sarkar2021phase,hahn2019quantum,fischer2021distributing}. Such a construction might be based on entanglement routing via passive optics for hypergraphs \cite{fainsin2025entanglement}. An approach in quantum metrology, as applied to graph states \cite{PhysRevLett.124.110502}, can be generalized to hypergraphs. Despite not being directly related to hypergraph states, another valid proposal relates to the generalization of quantum graphs to quantum hypergraphs \cite{Berkolaiko,PhysRevLett.79.4794}; In this case, the edges are no longer purely algebraic entities, but assume a metric in order to define a differential equation on it. The mathematical challenge remains in defining an appropriate metric over a hyperedge in such a way that it does not define a surface as usually done for quantum billiards problems \cite{fcwn-gnr2}.

Although the difficulties faced in the laboratory for many particles systems, practical implementations of hypergraph states in quantum computation are also a topic of research, including weighted hypergraphs for quantum databases using different quantum key distribution protocols and secret sharing, as well as quantum error correction \cite{PRA.78.042309.2008,balakuntala2017quantum}. Also, it is worthy to note that the states analyzed in \cite{miller2016hierarchy} are examples of hypergraph states, useful for a potential generalization of measurement-based quantum computation using only single-qubit Pauli measurements in Union Jack lattices. The resource states in this case are Union Jack states that generalize cluster states since the first possess $2D$ symmetry-protected topological order \cite{PhysRevB.80.155131}, in contrast to the $1D$ order of the cluster states. 

The non-trivial entanglement behaviour of mixed hypergraph states is also open to further investigation in more general cases, aiming to provide a definitive explanation of the non-trivial interplay among hyperedges of different cardinalities \cite{PhysRevA.109.012416}. The rapid development of controlled systems in practice indicates that the practical implementation of hypergraph states will be widespread, as was the case with graph states a decade ago \cite{lu2007experimental}, so we expect that implementations of hypergraph states shall grow as the optical devices technology advances,  such as the architectures for universal measurement-based and fusion-based quantum computing, known as successful architectures for graph states, and a promising topic for experimentalists \cite{PRL.86.5188.2001a,knill2001scheme,bartolucci2023fusion,thomas2024fusion}.

\section*{Aknowledgements}

V. Salem acknowledges the Spanish MCIN with funding from the European Union Next Generation EU (PRTRC17.I1) and Consejería de Educación from Junta de Castilla y León through the QCAYLE project and grant No. PID2023-148409NS-i00 MTM funded by AEI/10.13039/501100011033, and RED2022-134301-T.

\bibliographystyle{quantum}
\bibliography{bib.bib}

\end{document}